\documentclass{aip-cp}
\usepackage[numbers]{natbib}
\usepackage{rotating}
\usepackage{mathrsfs}
\usepackage{amssymb}
\usepackage[normalem]{ulem}
\usepackage{bm}

\usepackage{times}

\renewcommand{\rm}[1]{\textrm{#1}}

\newcommand{\beq}{\begin{equation}}
\newcommand{\eeq}{\end{equation}}
\newcommand{\calH}{ {\cal H} }

\newcommand{\rmi}{ {\rm i} }

\begin{document}

\title{Delineating the properties of matter in cold, dense QCD }
\author[aff1]{Toru Kojo}
\corresp[cor1]{Corresponding author and speaker: torujj@mail.ccnu.edu.cn}
\affil[aff1]{Key Laboratory of Quark and Lepton Physics (MOE) and Institute of Particle Physics, Central China Normal University, Wuhan 430079, China}

\maketitle

\begin{abstract}
The properties of dense QCD matter are delineated through the construction of equations of state which should be consistent with QCD calculations in the low and high density limits, nuclear laboratory experiments, and the neutron star observations. These constraints, together with the causality condition of the sound velocity, are used to develop the picture of hadron-quark continuity in which hadronic matter continuously transforms into quark matter (modulo small 1st order phase transitions). For hadronic matter (at baryon density $n_B \lesssim 2n_0$ with $n_0 \simeq 0.16\,$ fm$^{-3}$ being the nuclear saturation density) we use equations of state by Togashi et al. based on microscopic variational many-body calculations, and for quark matter ($n_B \gtrsim 5n_0$) we construct equations of state using a schematic quark model (with strangeness) whose interactions are motivated by the hadron phenomenology. The region between hadronic and quark matters ($2 n_0 \lesssim n_B \lesssim 5n_0$), which is most difficult to calculate, is treated by highly constrained interpolation between nuclear and quark matter equations of state. The resultant unified equation of state at zero temperature and $\beta$-equilibrium, which we call Quark-Hadron-Crossover (QHC18 and QHC19), is consistent with the measured properties of neutron stars and in addition gives us microscopic insights into the properties of dense QCD matter. In particular to $\sim 10n_0$ the gluons can remain as non-perturbative as in vacuum and the strangeness can be as abundant as up- and down-quarks at the core of two-solar mass neutron stars. Within our modeling the maximum mass is found less than $\simeq 2.35$ times solar mass and the baryon density at the core ranges in $\sim 5$-8$n_0$. 

\end{abstract}

\section{Introduction}\label{sec:intro} 

The determination of the QCD phase structure at large baryon density has been a difficult problem. Theoretically, this is partly because the lattice Monte Carlo simulations based on the QCD action suffer from the sign problem, and partly because nuclear many-body problems are very complex due to the strong short range correlations and the channel dependence of baryon-baryon interactions. Reliable theoretical calculations are possible only for dilute matter of hadrons or very dense matter of weakly coupled quarks and gluons. Inbetween the matter is supposed to be strongly correlated. Exploring the properties of such matter is the subject of this talk.

The calculations of strongly correlated matter is theoretically difficult but we can refer to recent advancements in astrophysical and laboratory observations for help. The astrophysical observations include: 
(i) the discoveries of two-solar mass ($2M_\odot$) neutron stars\footnote{After the submission of the original article, a neutron star with the mass $2.17^{+0.11}_{-0.10} M_\odot$ (at 68.3\% confirmation level) appeared \cite{2.17mass}. The estimate is based on the Shapiro delay technique as in \cite{fonseca}. The error will be reduced by the long term observation.} \cite{fonseca, Demorest,Antoniadis2013, 2.17mass},
the binary millisecond pulsar J1614-2230, with mass $1.928\pm0.017 M_\odot$ \cite{fonseca} (the original mass measurement was $1.97\pm  0.04 M_\odot$ \cite{Demorest}), and the pulsar J0348+0432 with mass $2.01 \pm 0.04 M_\odot$ \cite{Antoniadis2013}; (ii) 
X-ray analyses from neutron stars (reviewed in \cite{timing}) indicating the neutron star radii of $R=10$-$13$ km; (iii) the discovery of gravitational waves (GW170817 \cite{TheLIGOScientific:2017qsa,GW170817A}) from neutron star mergers as well as the electromagnetic counterparts \cite{Monitor:2017mdv,Coulter:2017wya}, which put the lower and upper bounds on the neutron star mass and radius. In addition there are important informations from supernovae and neutron star cooling but these topics will not be touched in this article. These astrophysical observations for 1-2$M_\odot$ stars are most directly correlated with equations of state at the baryon density, $n_B \gtrsim 2n_0$ ($n_0 \simeq 0.16\,$ fm$^{-3}$), while the domain of $n_B = 1$-$2n_0$ can be accessed from the laboratory nuclear experiments including the nuclear structure at $n_0$ \cite{Lattimer:2012xj,Oertel:2016bki} and heavy ion collisions \cite{Li:2008gp,Danielewicz:2002pu,Russotto:2011hq,Russotto:2016} in which collective flows and particle production are sensitive to equations of state used in transport calculations. We discuss how to implement these findings into our construction of equations of state.

Before directly approaching phenomenologically most relevant region, first we will outline the overall picture for cold, dense QCD matter from hadronic to quark matter region. Next we consider the empirical constraints on equations of state in the intermediate density regime where theoretical descriptions are most uncertain. Through attempts to make equations of state compatible with these constraints, we are lead to the hadron-quark continuity picture with which we consider the physics of strongly correlated quark matter and its connection to the hadronic matter beyond the purely nucleonic regime. More details about the theory can be found in our review article \cite{Baym:2017whm} and in a recent paper \cite{Baym:2019iky}. The tables for the equations of state are available on CompOSE (Compstar) repository, QHC18 \cite{Baym:2017whm} at [https://compose.obspm.fr/eos/139/], and QHC19 \cite{Baym:2019iky} at [https://compose.obspm.fr/eos/140/]. In this article we take the natural units $c=\hbar=1$ for the light velocity and the Planck constant.

\section{Outline: the picture of dense QCD matter from hadrons to quarks}\label{sec:outline} 

\begin{figure}[tb]

\centering
\includegraphics[scale=0.32,clip]{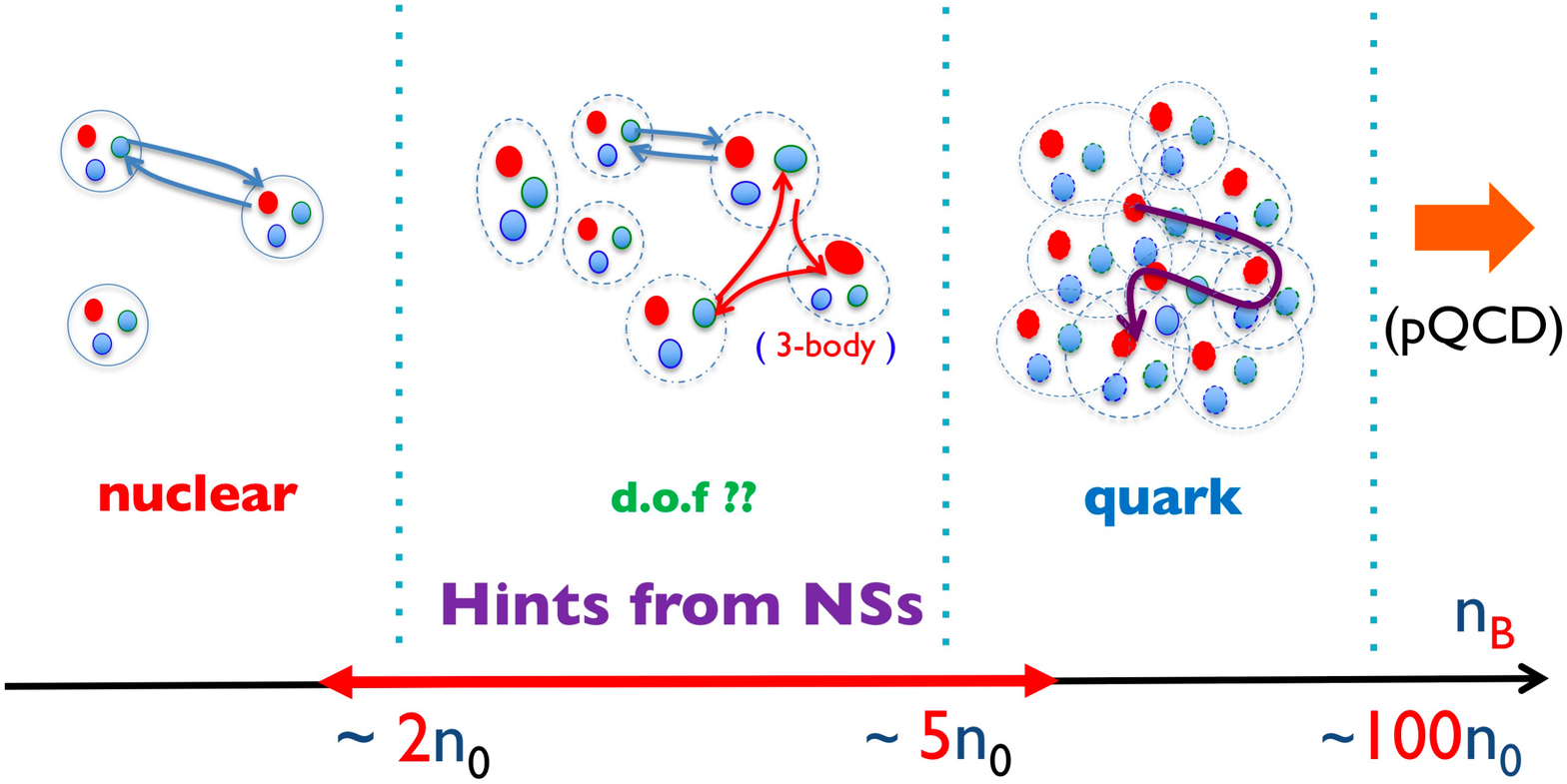} 
\caption{
The three window description.
  }
\vspace{-2.5cm}
   \label{fig:3window}
\end{figure}

To outline, let us first classify various regimes with the range of baryon density. These regimes are specified by the validity range of microscopic calculations. At low enough density quarks and gluons are confined and the resultant hadrons should be good effective degrees of freedom for calculations of the matter properties. Near the saturation density the system is described as a liquid of nucleons interacting through two- and three-body forces. Those nuclear forces are fixed by nucleon-nucleon scattering data and the properties of light nuclei, and they are used as the inputs for the nuclear many-body calculations which must be non-perturbative to deal with the soft nucleonic excitations from the nucleon Fermi sea  \cite{Akmal:1998cf,Togashi:2017mjp}. The microscopic calculations reproduce the empirical saturation properties to 10-30\% accuracy. One of major uncertainties come from the microscopic three-body forces as well as the treatments of three-body correlations. At $n_B\simeq n_0$, the two-body contributions in equations of state dominate while the three-body contribution is important but the size is by itself small. But around $n_B \simeq 2n_0$ the overall size for two-body and three-body contributions become comparable (see for instance Table V in the paper by Akmal-Pandharipande-Ravenhall (APR) \cite{Akmal:1998cf}), indicating the importance of four-, five-, and more-body forces. These forces are due to the exchange of mesons which are more microscopically described as quark exchanges.  With this quark exchange picture, it is natural to expect the structural changes of hadrons associated with the increasingly important many-body forces. 

At very high density the equations of state should be described by weakly coupled quarks and gluons, as the typical processes involve hard momentum transfer. The soft momentum transfer processes with strong coupling are less typical and hence suppressed by the phase space factor. The perturbative QCD calculations for equations of state have been done to the three-loop order with the resummation of particular kind of graphs \cite{Kurkela:2009gj,Fraga:2001id,Freedman:1976ub}. They show that the calculations do not show good convergence in the expansion of $\alpha_s$ below $\mu_B \lesssim 3$ GeV or $n_B \lesssim 50$-$100n_0$. This suggests the importance of interactions which are no longer quite perturbative below $n_B \lesssim 50$-$100n_0$. The matter is supposed to be strongly correlated quark matter.

The picture of the strong correlation for $\mu_B=1$-$3$ GeV is also supported from QC$_2$D which is the two-color version of QCD. In this theory there is no sign problem so that the lattice MonteCarlo simulations can be performed. Using relatively heavy current quark masses, seminal works studied the phase structure, equations of state, diquark condensates, Polyakov loops and so on \cite{Hands:2010gd,Cotter:2012mb,Braguta:2016cpw}. Also the Landau gauge gluon propagators and vertices have been measured \cite{Hajizadeh:2017ewa,Boz:2018crd}. Here we mention the difference and similarity between two-color and three-color QCD. They are very different in their dilute regimes because how to make hadrons is different. In QC$_2$D the color-antisymmetric wavefunction is color-singlet and the baryons are bosons. In addition, their masses are degenerated with those of the Goldstone bosons of the chiral symmetry breaking due to the Pauli-Gursey symmetry. As a consequence the baryonic matter in QC$_2$D starts with the Bose-Einstein condensate (BEC) phase of baryons. As the system enters the dense regime, however, the Pauli blocking effects acting on quarks inside of baryons anyway establish the quark Fermi sea, and the BEC changes into the BCS phase where the pairing is dominated by quarks near the Fermi surface. For the density high enough for the local color neutrality to be achieved, presumably the difference between two-color and three-color QCD is not so dramatic. With these qualifications in mind, it is worth mentioning that the lattice study for QC$_2$D found that the critical temperature of the BCS phase is $T_c^{BCS} = 80$-$120$ MeV which is almost constant in the quark matter regime to $\mu_B \simeq 3$ GeV. If the conventional relation $T_c = 0.57 \Delta$ holds for the gap $\Delta$, at the Fermi surface the quarks open the gaps of $175$-$210$ MeV, the natural non-perturbative scale in QCD. In addition the study of the gluon propagator in Landau gauge found that the gluon propagators are insensitive to $\mu_B$ from the vacuum to the dense quark matter regime to $\mu_q^{2-color} = \mu_B/2 \simeq 1$ GeV \cite{Boz:2018crd}. These findings support the idea of strongly correlated quark matter for $\mu_B=1$-$3$ GeV in three-color QCD.

Now we assemble the preceding discussions on the microscopic nature of matter. We consider the matter by decomposing it into three windows \cite{Masuda:2012kf,Masuda:2012ed,Kojo:2014rca,Kojo:2015fua}; the nuclear regime at $n_B \le 2n_0$; the quark matter regime at $n_B \ge 5n_0$; and the intermediate regime for $2$-$5n_0$. The picture we have is illustrated in Fig.\ref{fig:3window}. At low density, $n_B \le 2n_0$, the matter is dilute and baryons remain well-defined objects, so the equations of state are described by nuclear ones. Beyond $\sim 2n_0$ and to $\sim 5n_0$, it is unlikely that nucleons remain effective degrees of freedom but the density is not high enough for quarks to get deconfined from baryons. At $n_B \sim 5n_0$, baryons with the radii of $\sim 0.5$ fm start to touch one another, and one may expect quarks to travel among different baryons. This is where we expect the formation of the quark Fermi sea. In three-flavor matter the quark Fermi momentum is $p_F \sim 400$ MeV (for two-flavor matter $p_F$ is even larger).

Theoretically the most uncertain is the domain of $n_B=2$-$5n_0$. The difficulty is largely due to the confining effects in QCD which transform effective degrees of freedom. Meanwhile phenomenologically this domain is most important in the physics of neutron stars. Fortunately the observations in the neutron star physics have improved significantly and now we have strong constraints on the properties of strongly correlated matter. For this reason we briefly review the constraints from recent neutron star observations, in particular the GW170817 event \cite{TheLIGOScientific:2017qsa,GW170817A} and associated electro-magnetic counterparts \cite{Monitor:2017mdv,Coulter:2017wya}, and then discuss the characteristic features relevant for equations of state. Also we take into account the constraints from the nuclear experiments at zero temperature and near the saturation density as well as heavy ion collisions at finite temperature and supra saturation density which are continuously improving our knowledges about the domain of $1$-$2n_0$.

\section{The  mass-radius relation and its implication for the QCD equations of state}

\begin{figure}[tb]
\vspace{0.0cm}
\centering
\includegraphics[scale=0.32]{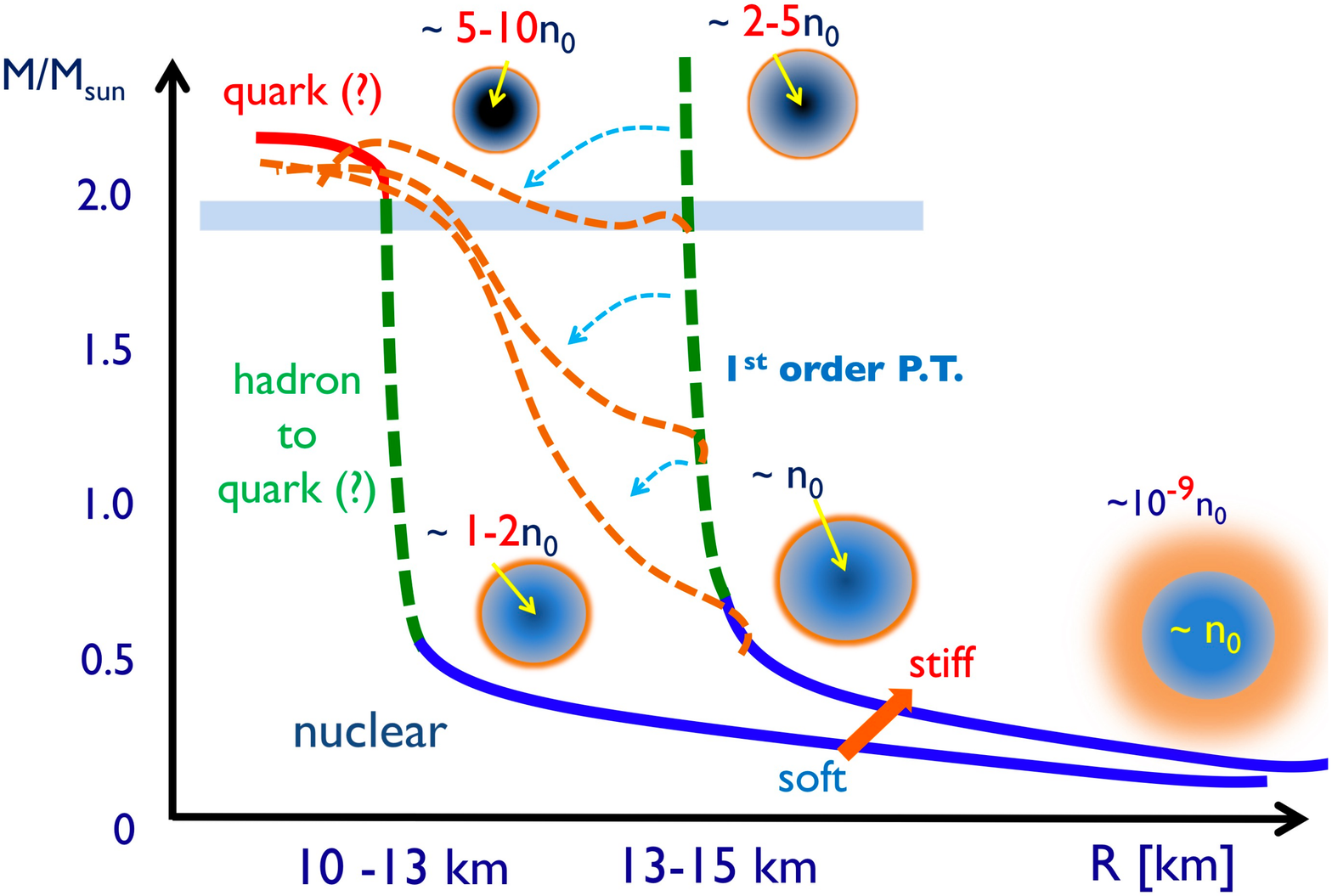}
\caption{
The correlation between the $M$-$R$ relation and equations of state.
  }
  \vspace{-0.5cm}
   \label{fig:M-R}
\end{figure}

One of the most fundamental quantities of neutron stars are their mass-radius ($M$-$R$) relation. They can be calculated by solving the Einstein equation coupled to QCD (+electroweak) energy-momentum tensor or the equation of state. If a star is not spinning too rapidly one can use the Tolman-Oppenheimer-Volkoff (TOV) equation which is based on the spherical symmetry. Equations of state enter the TOV equation in the form of $P(\varepsilon)$, where $P$ is pressure and $\varepsilon$ is energy density. The $M$-$R$ relation and the equation of state has one-to-one correspondence \cite{Lindblom1992} and in principle one can determine the QCD equation of state directly from observed $M$-$R$ relations. In reality the measured $M$-$R$ relations have errors, and we need the inversion process with weighted probability, see for instance \cite{Fujimoto:2017cdo,Fujimoto:2019hxv}.

In order to characterize the structure of neutron stars we usually discuss the stiffness of equations of state. First we begin with some terminology defined for this article. ``Stiff" equations of state are those with large pressure $P$ at given energy density $\varepsilon$. The stiffer equations of state generally lead to larger maximum masses and larger radii for neutron stars; the energy density attracts the material to the center while the pressure increased in the compressed matter prevents the matter to collapse. We will not use the speed of sound $c_s=(\partial P/\partial \varepsilon)^{1/2}$ as the measure of the stiffness, because the derivative does not specify where $P(\varepsilon)$ starts. Indeed, if we start with a very stiff initial condition for $P(\varepsilon)$, even ideal gas equations of state with $c_s^2=1/3$ can generate very large maximum masses. This happens, e.g. for a bag model for ideal gas quark matter with a small bag constant, $P= P_{ ideal } - B$, which produces the massive neutron star with the mass $\ge 2M_\odot$ when $B$ is $\lesssim 56$ MeV fm$^{-3}$. This example already shows how the important nuclear physics is to constrain possible scenarios. (Below we will not discuss absolutely stable quark star scenarios in which nuclear matter is regarded as a metastable state \cite{Witten:1984rs}.)

Secondly we should specify at which region of density the equations of state are stiff. As we will see this is crucial to classify equations of state and modern observational constraints. We will use the terminology ``soft-stiff", by which we mean that equations of state is soft at low density, $n_B \le 2n_0$, and stiff at high density, $n_B \ge 5n_0$. For the reasons described below, equations of state leading to $R_{1.4} \le 13$ km for $1.4M_{\odot}$ stars will be called ``soft at low density", and equations of state leading to $M \ge 2M_{\odot}$ will be called ``stiff at high density". Then the soft-stiff equations of state generate the $M$-$R$ curves with the typical radii of $R_{1.4} \le 13$ km and the maximum mass $\ge 2M_\odot$. The terminology of the other combinations, such as ``stiff-stiff", ``stiff-soft", etc., should be already clear from this explanation.

It has been known \cite{Lattimer:2006xb}  that the shape of a $M$-$R$ curve has strong correlations with equations of state at several fiducial densities, see Fig.\ref{fig:M-R}. For a light star with very low baryon density at the core, the material outside of the core is loosely bound by the gravity, so the corresponding radius, defined at the point of $P=0$, is very large. For a star with slightly larger core density, the neutron star mass $M$ is slightly larger but the core $R$ is significantly smaller, because even slightly increased gravity can easily compress the loosely bound material. This rapid change in $R$ for a $M$-$R$ curve continues until the loosely bound material becomes very thin, and then we observe the radius of a dense matter which is dominated by a nuclear liquid at $n_B =1$-$2n_0$ where the repulsive forces are important. As a result the matter is no longer compressed substantially and this determines the overall size of neutron star radius. If a $M$-$R$ curve enters this regime the curve goes vertically. Eventually the curve reaches the maximum in $M$ at $n_B \gtrsim 5n_0$. Due to the existence of two-solar mass ($2M_\odot$) neutron stars \cite{Demorest:2010bx,Antoniadis:2013pzd}, equations of state at high density must be stiff otherwise they lead to the collapse of the stars before the mass reaches $2M_\odot$. 

At this point it should be mentioned how the characteristics of equations of state are reflected in the shape of $M$-$R$ curves. For this purpose we compare the soft-stiff equations of state with the stiff-stiff ones. The soft-stiff combination leads to small radii by soft low density equations of state while the maximum mass is large; the resulting stars are very compact and baryon density tends to be high. This type of equations of state is hard to accommodate the first order phase transition in the domain $2$-$5n_0$; starting with soft equations of state,  the first order phase transition and the associated softening easily makes neutron stars unstable to the gravitational collapse. In contrast, the stiff-stiff type of equations of state can remain reasonably stiff even after the first order transition, and some of them can pass the $2M_\odot$ constraint. After the first order transition, the radius in a $M$-$R$ curve shrinks rapidly and approaches the curves of soft-stiff equations of state. In extreme case the equations of state lead to the third family of stars in which the $M$-$R$ curves jump to another branch of curves \cite{Maslov:2018ghi,Benic:2014jia}. The recent studies for the signatures of possible first order transition, see Refs.\cite{Most:2018eaw,Bauswein:2018bma,Han:2018mtj} for neutron star mergers, and Ref.\cite{Fischer:2017lag} for supernovae explosion of blue supergiant stars; these studies tell us how loud the signatures can be.

\section{The observational constraints }

\begin{figure}[tb]
\vspace{0.0cm}
\centering
\includegraphics[scale=0.35]{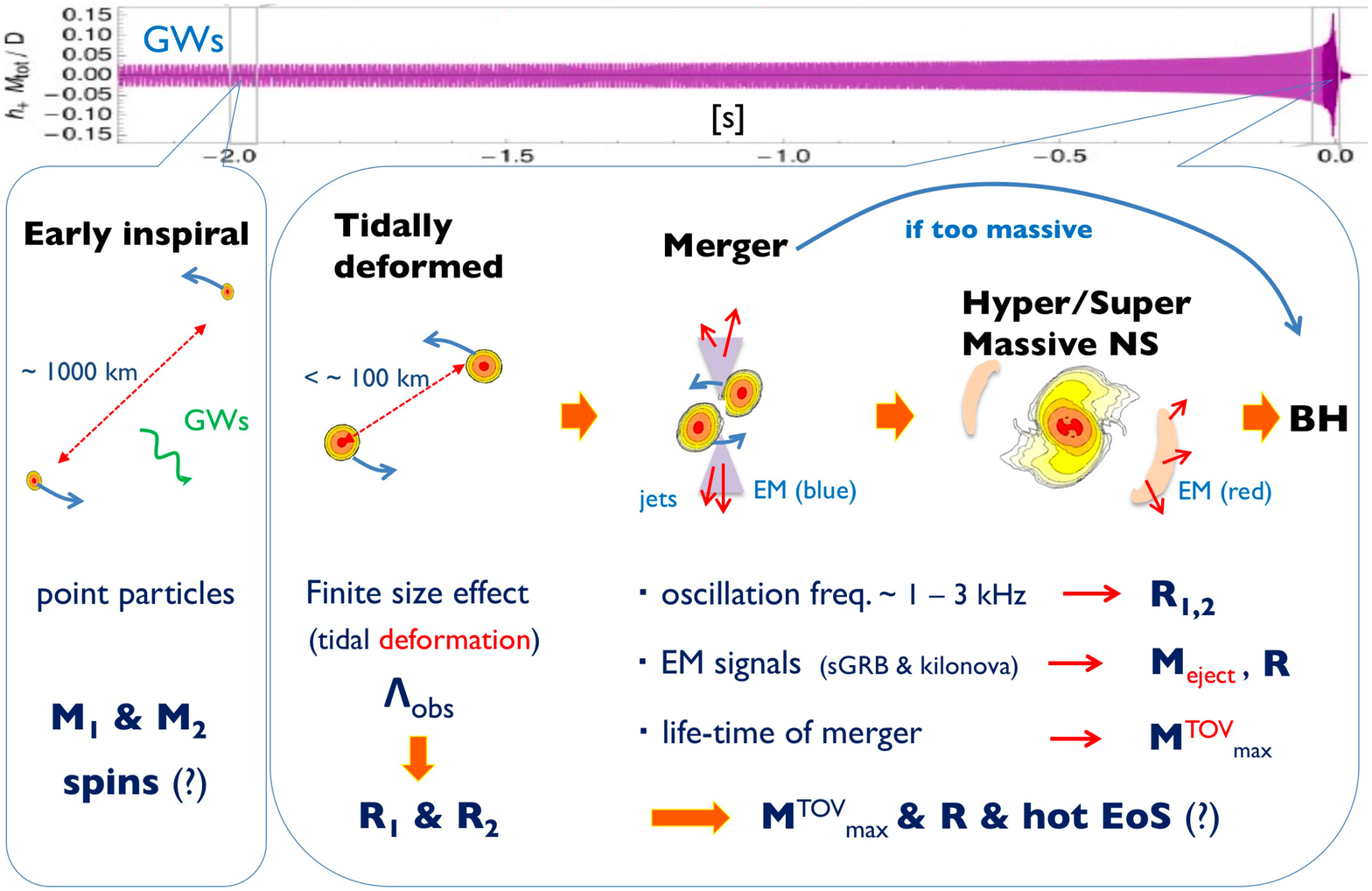}
\caption{
The time evolution of a neutron star merger.
  }
  \vspace{-0.5cm}
   \label{fig:NSmergers}
\end{figure}

The difference between the soft-stiff and stiff-stiff type equations of state can be found in neutron star radii and nuclear laboratory experiments at low energy. 

The observational constraints on $R_{1.4}$, which have been based on spectroscopic analyses of the X-rays from the neutron star surface, contain several systematic uncertainties including the distance to the source, the atmospheric composition, the assumption of black-body radiation, etc. The current trend, however, is converging toward the estimate $R_{1.4}=10$-$13$ km \cite{timing}. The X-ray analyses will be improved in NICER (the Neutron Star Interior Composition Explorer) experiment \cite{nicer,michi,Miller:2016kae,ozel-nicer} on the International Space Station. The NICER was launched in 2017. The NICER will remove a number of systematic uncertainties by performing the phase-resolved spectroscopy for X-rays emitted from hot spots on the neutron star surface which show periodic motion associated with the rotation of stars. It will determine $R_{1.4}$ to $\simeq 1$ km accuracy.

The equations of state at  $1$-$2n_0$ can be also inferred from the study of nuclear symmetry energy, $J$, and its slope parameter, $L$, at the saturation density which appears as the coefficients of the density expansion around $n_0$,
\beq
\bar{E} (n_B, x_p=0) - \bar{E} (n_B,  x_p=0.5) = J + L \frac{\, n_B-n_0 \,}{\, 3 n_0 \,} + \cdots \,,
 \eeq
where $\bar{E}$ is the energy per particle and $x_p$ is the proton fraction. The parameter $L$ appears in the pressure in pure neutron matter at $n_0$ as $P(n_0, 0) = L n_0 /3$ and hence are important parameters for the $M$-$R$ relation in the low mass region. The results of nuclear laboratory experiments for the neutron skin and giant monopole resonance as well as the results of many-body calculations with nuclear forces calculated from chiral effective theories are summarized in \cite{Lattimer:2012xj}. The weighted average leads to $J = 31.6$ MeV and $L = 58.9$ MeV with the error margin $\Delta J = 2.7$ MeV and $\Delta L = 16$ MeV. More recent estimate by another group leads to $J = 31.7\pm 3.2$ MeV and $L = 58.7 \pm 28.1$ MeV \cite{Oertel:2016bki}. Unless something drastic happens for higher order terms in powers of $(n_B-n_0)$, the extrapolation of equations of state to $\sim 2n_0$ leads to $R_{1.4}=10$-$13$ km.

Meanwhile independent constraints for the $2$-$5n_0$ domain come from heavy ion experiments \cite{Li:2008gp}. Heavy ions experiments are most directly related to the properties of symmetric nuclear matter. The study of collective flows from heavy ion experiments at incident energy up to 10 MeV/nucleon, with the aid of transport models including neutron, proton, delta, and pions, has estimated the pressure up to $\sim 4n_0$ \cite{Danielewicz:2002pu}. More recently, equations of state for asymmetric matter to $\sim 2n_0$ is estimated by analyzing the differential flows for neutrons and charged particles from $^{197}$Au-$^{197}$Au collisions with the incident energy 400 MeV/nucleon in the FOPI-LAND experiment \cite{Russotto:2011hq}. The experiment ASY-EOS further improved the precision \cite{Russotto:2016}. The current estimate suggests that stiff equations of state are not compatible with the data, and they are consistent with soft equations of state leading to neutron star radius $R_{1.4}=10$-$13$ km.

All these results seem to suggest rather soft equations of state leading to $R_{1.4} <13$ km. But the X-ray analyses and heavy ion experiments, relevant for $\gtrsim 2n_0$, still contain systematic uncertainties associated with the modeling, and further independent analyses based on different assumptions are called for. Such analyses in turn will also clarify the validity of the assumptions used in the previous analyses.

In this respect the neutron star merger events, one of which was discovered in 2018 for the first time (with the gravitational waves, GW170817 \cite{TheLIGOScientific:2017qsa,GW170817A}, the gamma-ray burst, GRB170817A  \cite{Monitor:2017mdv}, and the kilo-nova signals, SSS17a/AT 2017gfo \cite{Coulter:2017wya}), are extremely valuable sources for static and dynamical aspects of neutron stars. In next 10 years much more events of the order hundred are expected to come, see a white paper \cite{Foley:2019evo}. In April 2019, the the LIGO-Virgo collaboration will start its third observing run, O3, to March 2020, and KAGRA may join the network toward the end of the O3 operation. 

\begin{figure}[tb]
\vspace{0.0cm}
\centering
\includegraphics[scale=0.35]{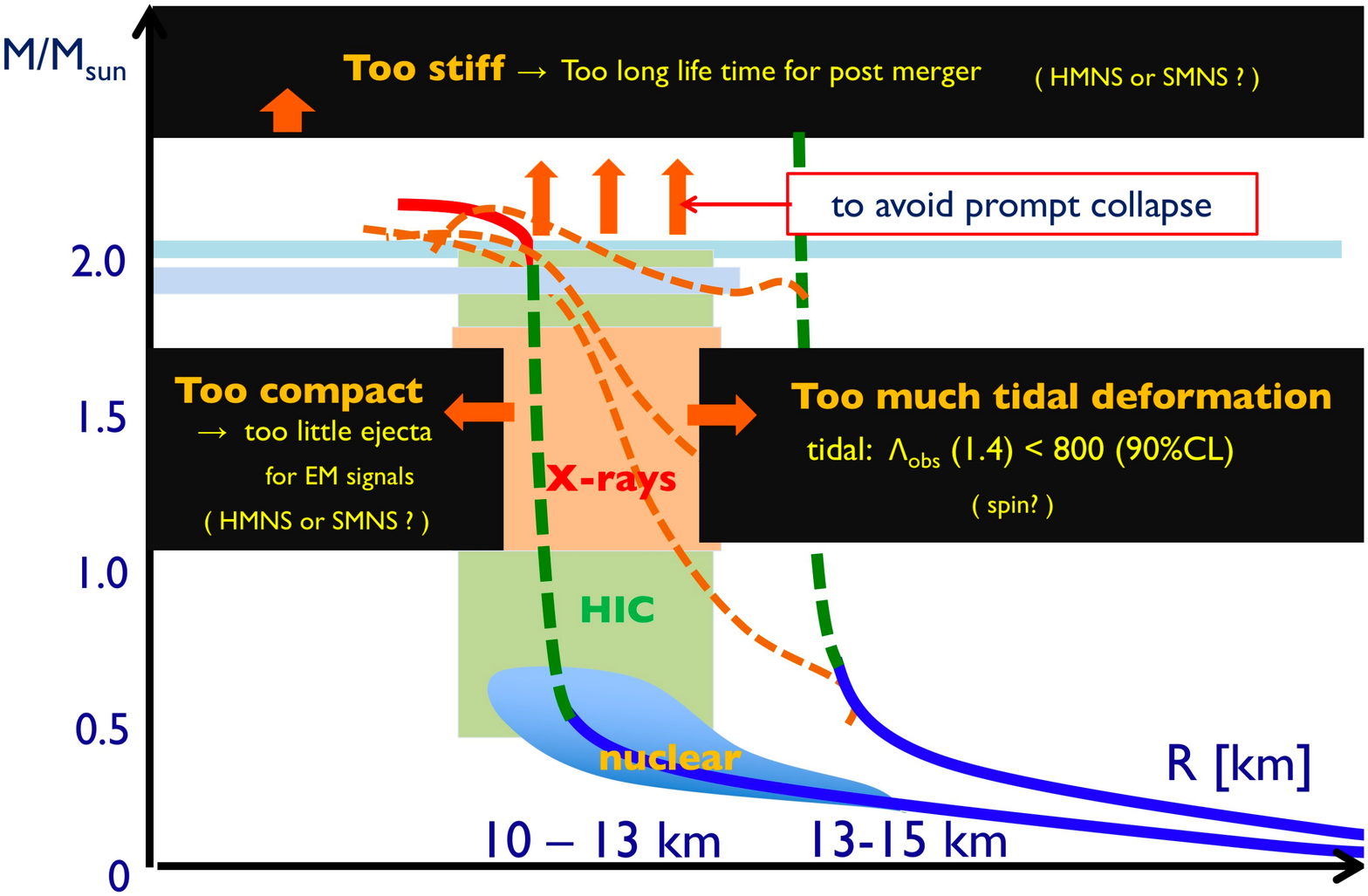}
\caption{
A rough sketch of various constraints in the $M$-$R$ curves.
  }
  \vspace{-0.5cm}
   \label{fig:constraints}
\end{figure}

They contain several different stages in the evolution, see Fig. \ref{fig:NSmergers}. We first outline the idea of how to obtain constraints and then comment on the values given in seminal works.

Initially the two neutron stars are widely separated and they can be treated as point particles. The spiraling motion depends primarily on the masses, and spins may enter as the higher order in the velocity expansion. The chirp mass, ${\cal M} =(M_1 M_2)^{3/5}/(M_1+M_2)^{2/5}$, as a combination of mass of each star is precisely determined, $1.188^{+0.004}_{-0.002} M_\odot$, which leads to the total mass,  $M^{ GW170817} =2.73$-$2.78 M_\odot$ for plausible mass ratio $M_1/M_2$ of $0.7$-$1.0$.

They emit gravitational waves and after long time they come close each other. At distance of $\sim 100$ km, the finite size effects become important; a star feels the tidal field from the other star and it causes the tidal deformation in each star. Such deformation in effect adds to the attractive force and hence accelerates the merging process. An important quantity in this discussion is the dimensionless tidal deformability, $\Lambda = \frac{2}{3} k_2 (R/G_N M)^5$ ($G_N$: Newton constant; $k_2$: Love number \cite{Hinderer:2015}), of each star before the coalescence. Clearly $\Lambda$ is very sensitive to the compactness and radius of the star; smaller $\Lambda$ for more compact stars. At leading order, the tidal deformability of neutron stars 1 and 2 enter the gravitational waveform in the combination
\beq
 \tilde{\Lambda} = \frac{16}{13}\frac{\, (M_1+12M_2)M_1^4\Lambda_1 + (12M_1+M_2)M_2^4\Lambda_2}{ (M_1 + M_2)^5 } \,,
\eeq
which depends on the total mass and the mass ratio between $M_1$ and $M_2$. As far as stars are separated, the tidal effects may be taken into account as perturbations while their information is accumulated for a long time before the merger, thus they can be regarded as a clean signal with detectable strength; this is the frequency domain where the advanced LIGO-Virgo collaboration to O2 was able to detect in the GW170817 event. The obtained is the upper bound on $\Lambda$ and in turn on the radius for a star with $\simeq 1.4M_\odot$.

 When two neutron stars merge, the system enters the highly nonlinear regime, producing gravitational waves at high frequency of 1-3 kHz. The frequency strongly depends on the compactness of merged stars, and more compact stars produce waves at higher frequency. In this nonlinear regime the physics is very rich. Unfortunately right now the gravitational waves from this frequency domain were not detected by the detector sensitivity achieved in O2, but will be measurable after the detector upgrade. For this reason the fate of the merger is not fully conclusive, as we have not measured the collapse time. The merger promptly collapses to a black hole if the merger is too massive, while the merger survives for a while as a metastable state, either in the forms of hypermassive or supramassive neutron stars which supports materials by differential or rigid rotations, respectively, until the angular momenta and the magnetic fields are dissipated. A hypermassive neutron star survives for 10-100 ms if the mass is less than $\sim 1.5M^{max}_{TOV}$ ($M^{ max }_{ TOV }$: the maximum mass for a non-rotating star), while a supramassive star survives the collapse for longer time scale of $\sim 100$ s if the mass is less than $\simeq 1.2M^{max}_{TOV}$. Therefore the measurement of the lifetime of the merger, together with the estimated remnant mass, can constrain the range of $M^{ max }_{ TOV }$; remarkably it can constrain not only the lower bound given by the existence of the neutron stars, but also the upper bound by the non-existence of supramassive neutron stars (see below for more concrete examples).

Although the gravitational waves from post mergers were not detected, there are ample electromagnetic signals including the gamma-ray burst and kilonova. The luminosity of electromagnetic signals are sensitive to the ejecta from the merger process. The high mass ejecta of $\sim 10^{-1}$-$10^{-2} M_\odot$ was inferred from electromagnetic transient. The prompt collapse seems incompatible with the observed signals as they do not have much time to produce sufficient amount of the ejecta. Similarly the coalescence of too compact stars does not produce much ejecta either. This places the lower bound on the neutron star radius. Meanwhile whether the merger remnant was hypermassive and supramassive neutron stars is not fully clarified by the electromagnetic signals. Naturally it affects estimates of $M^{max}_{TOV}$.

Shown in Fig.\ref{fig:constraints} is a rough sketch of the constraints discussed so far (some of them are translated into the $M$-$R$ relation). Now we quote some of estimates available in the literature. 

The maximum mass should exceed the established lower bounds $1.928\pm0.017 M_\odot $ \cite{fonseca} or $1.97\pm  0.04 M_\odot$ \cite{Demorest}. The constraints from GW170817 depend on the hypermassive and supramassive neutron star scenarios. In the former scenario, the constraints for the upperbound were obtained; $\lesssim 2.17 M_\odot$ (Margalit and Metzger \cite{Margalit:2017dij}); $\lesssim 2.16^{+0.17}_{-0.15} M_\odot$ (Rezzola et al. \cite{Rezzolla:2017aly}); $\lesssim 2.16 M_\odot$ (Ruiz et al. \cite{Ruiz:2017due}). Meanwhile another group favors a long-lived massive neutron stars with a torus to have a strong neutrino emitter, leading to higher mass, $2.15$-$2.25 M_\odot$ (Shibata et al. \cite{Shibata:2017xdx}). Another group also favors the long-lived supramassive star scenario to find a source to explain the late emission in the kilonova event (Yu et al. \cite{Yu:2017syg}); this leads to $\gtrsim (2.73$-$2.78M_{\odot}$)$/1.2 \simeq 2.28$-$2.32 M_\odot$ from the condition $M^{ GW170817} \gtrsim 1.2M_{TOV}$. The diversity of these estimates will be settled by the direct measurement of the collapse time signaled by the gravitational waves at high frequency.

The lower bound of the neutron star tidal deformability, which is strongly correlated with the radius, was given by the condition for sufficient amount of ejecta consistent with the electromagnetic observations. Bauswein et al. \cite{Bauswein:2017vtn} gave the constraint $R_{1.6} \ge 10.68^{+0.15}_{-0.04}$ km. Based on the similar reasoning Radice et al. \cite{Radice:2017lry} gave the constraint $\tilde{\Lambda}_{1.4} \gtrsim 400 $ which in turn requires $R_{1.4} \gtrsim 11.0$-$11.5$ km excluding many equations of state, e.g., the APR. We note that the constraint on the lower bound was based on the post merger dynamics for which systematic errors are not fully understood. Meanwhile the upper bound is derived from tidal deformed phase in which the system is cleaner. The upper bound of $\tilde{\Lambda}$ was directly estimated from the analyses of gravitational waves as $\Lambda_{1.4} \lesssim 800$ (CL90\%), and it is translated to the radius constraint $R_{1.4} \lesssim 13.6$ km. Useful summary in terms of $M$-$R$ curves can be found in \cite{Bauswein:2017vtn,Annala:2017llu}.

Taking all these constraints together we believe it reasonable to conclude that the equations of state for neutron stars are the soft-stiff type. Next we will discuss the implications from this supposed equations of state.

\section{Soft-stiff equations of state and quark-hadron continuity }

\begin{figure}[tb]
\vspace{0.0cm}
\centering
\includegraphics[scale=0.3]{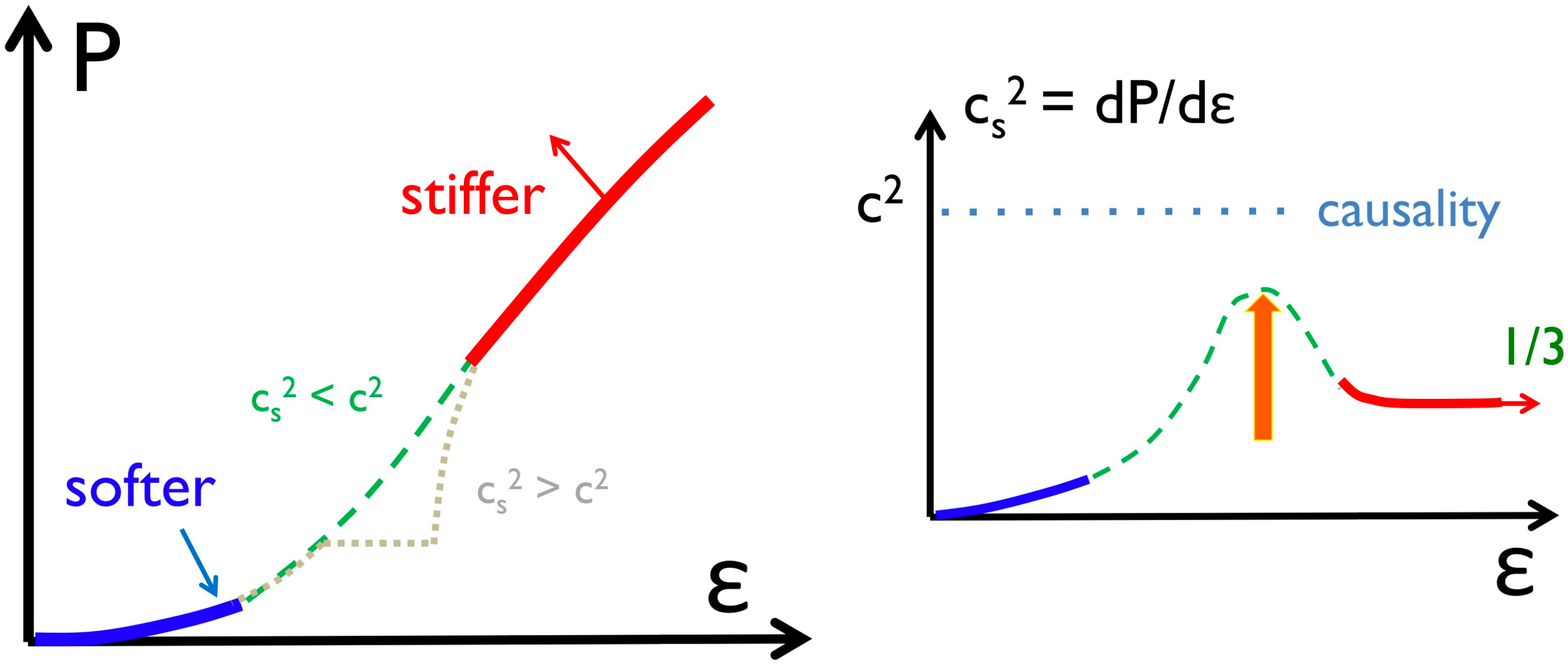}
\caption{
The pressure vs energy density and speed of sound with or without the first order phase transitions.
  }
  \vspace{-0.5cm}
   \label{fig:cs2}
\end{figure}

The soft-stiff equations of state have rather peculiar structure (Fig.\ref{fig:cs2}). In order to connect soft equations at low density to stiff ones at high density, there must be a domain where the pressure grows rapidly as a function of energy density. This means that the speed of sound (square), $c_s^2=\partial P/\partial \varepsilon$, become large for the domain $2$-$5n_0$. But assuming the causality condition, the speed of sound must be smaller than the light velocity. This constrains the structure of $P(\varepsilon)$ curves, and the condition becomes severer for softer-stiffer equations of state. In the context of the QCD phase structure, the strong first order phase transitions are disfavored as it significantly softens equations of state; at the phase transition point the pressure stays constant but the energy density jumps, leading to $c_s^2=0$. In contrast the soft-stiff equations of state demand the appearance of peaks in $c_s^2$ exceeding the conformal limit $1/3$, provided that the equations of state eventually approach the pQCD result with $c_s^2 \simeq 1/3$ at high density. If there are first order phase transitions there must be even stronger peaks nearby the phase transition points; otherwise  the softening induced by the first order phase transitions cannot be compensated. For more systematic arguments we refer to \cite{Alford:2013aca,Tews:2018kmu}.

This situation leads us to the picture of quark-hadron continuity in which hadronic matter continuously transforms into quark matter, without experiencing sharp thermodynamic phase transitions between distinct phases. 

Such continuity picture was developed in the context of the crossover from the superfluid hadronic phase with hyperons to the color-flavor-locked superconducting phase \cite{Schafer:1998ef}. In this argument the quantum numbers carried by hadronic degrees of freedom find their counterparts in quark matter. This scenario was revisited with questions concerning the dynamics \cite{Hatsuda:2006ps,Zhang:2008wx}, where the $U_A(1)$ anomaly, or more concretely the interplay between the chiral and diquark condensates, plays the key role. These studies are based on theoretical considerations and model calculations. In contrast we have reached the continuity picture in attempts to understand the neutron star constraints in a way consistent with the nuclear and hadron phenomenology. 

Our way of using the terminology ``continuity" is somewhat looser than the previous theoretical studies; right now the strict connection between order parameters in hadronic and quark matter is not our primary concern, as far as these order parameters do not change the thermodynamics of the system drastically. For instance the appearance of gaps of nuclear scale $1$-$10$ MeV in superfluid phases is, for the moment, not crucial ingredients in our arguments.  They will be more critical when we proceed to the discussion of the neutron star cooling. 

On the other hand we do care the connection to hadron phenomenology at QCD scale $\Lambda_{QCD} \sim 200$ MeV appearing in the hadron mass splitting and nuclear forces at short distance. We will come back to this point when we introduce a schematic quark model.

\section{The three-window modeling: general remarks}

\begin{figure}[tb]
\vspace{0.0cm}
\centering
\includegraphics[scale=0.25]{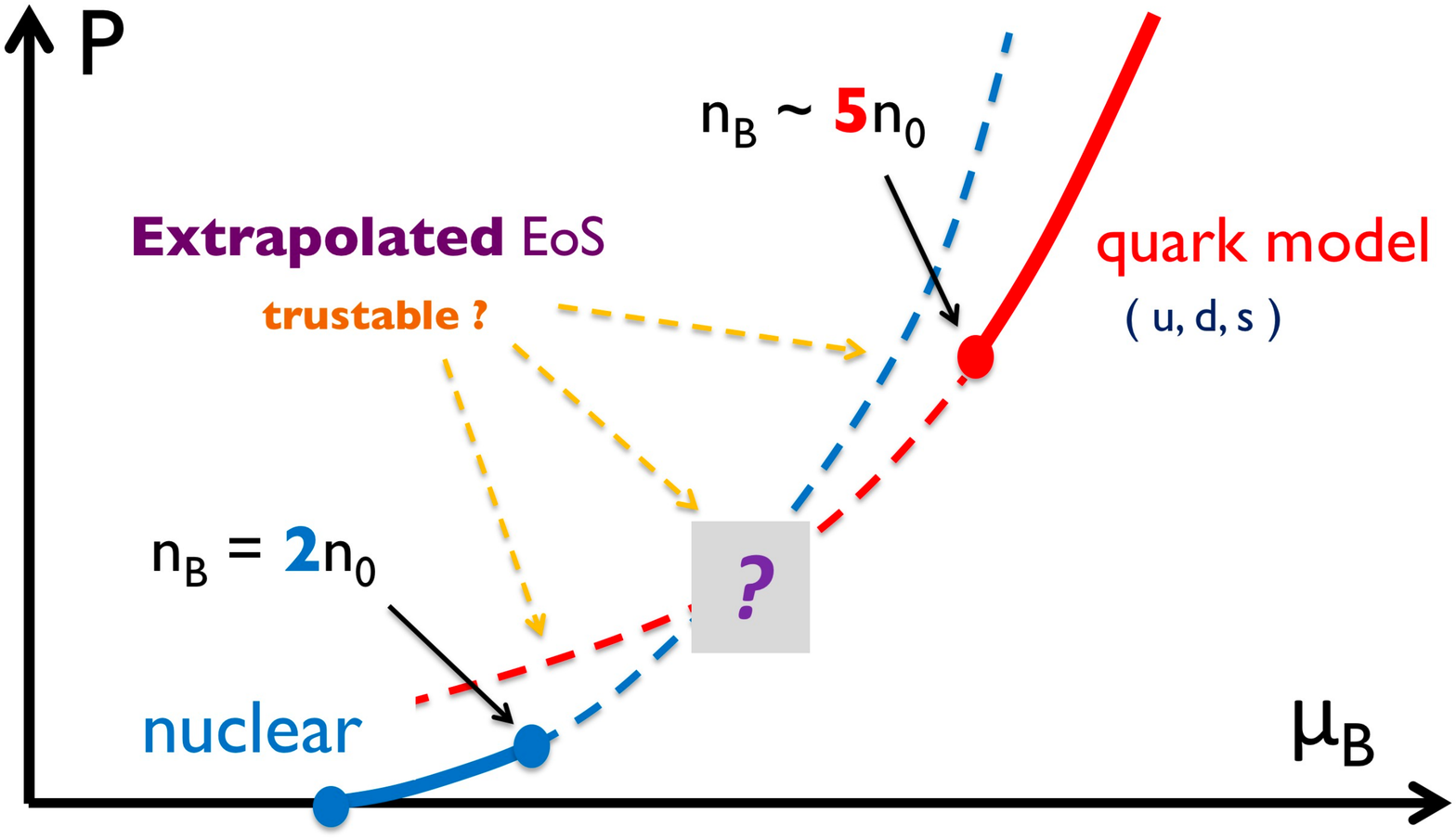}
\hspace{0cm}
\includegraphics[scale=0.25]{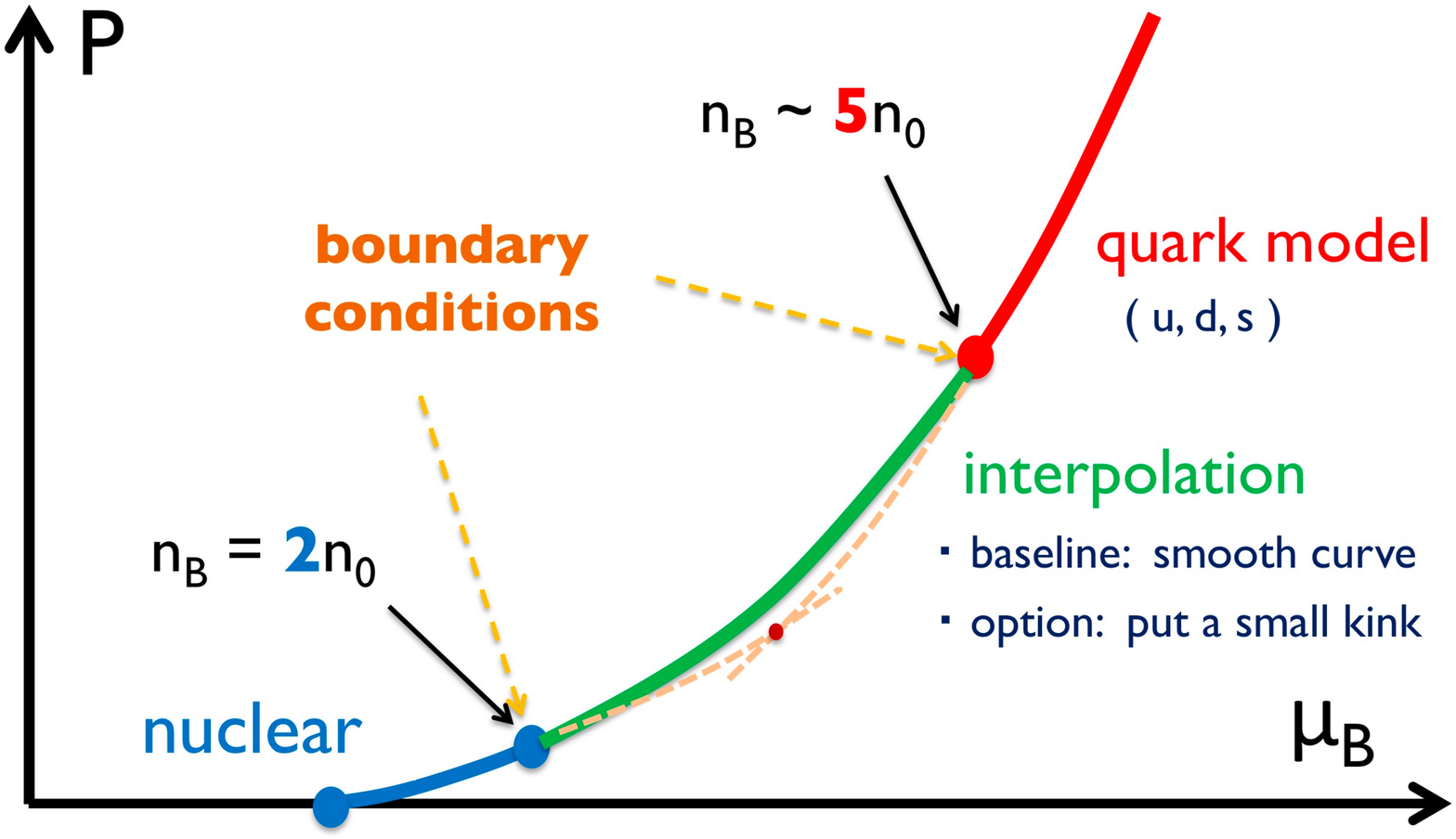}
\caption{
The hybrid equations of state in conventional (left) and three-window modeling (right). In the conventional modeling the quark pressure adopted in the figure is not acceptable while in the three modeling is used as a physical candidate supplying the boundary condition for the interpolated curve.  
  }
  \vspace{-0.5cm}
   \label{fig:hybrid}
\end{figure}

Based on the quark-hadron continuity picture we will construct equations of state within three-window modeling. The physical picture was already described (see Fig.\ref{fig:3window}) so now we turn to the practical use (Fig.\ref{fig:hybrid}). 

First of all we need to choose which functions to be interpolated for which variables. The best combinations are either $P(\mu_B)$ or $\varepsilon (n_B)$, from which all the other thermodynamic quantities can be derived. If we were to use $P(\varepsilon)$, we cannot derive $n_B$ from this function and we need to introduce another interpolation for $n_B$. Two independent interpolating functions are very likely incompatible with the thermodynamic consistency and should be avoided. We choose $P(\mu_B)$ throughtout.

We use a nuclear equation of state to $n_B \sim 2n_0$ but beyond which stop using it as the systematic uncertainties are uncontrollable. At high density we use a quark equation of state but stop using it below $\sim 5n_0$ as the confining effects, whose descriptions were not established, should play crucial roles. Then we interpolate these equations of state with polynomials \cite{Kojo:2014rca}
\beq
P(\mu_B) = \sum_{n=0}^5 c_n \mu_B^n \,.
\eeq
Clearly the form chosen is just one choice of interpolating functions. If we wish, we can use pressure curves including small kinks to describe the first order phase transitions, but the effects of kinks should be small according to the constraints discussed in the previous section. Therefore smooth curves are taken to be our baseline. To determine the coefficients $c_n$'s, we first compute $n_B=\partial P/\partial \mu_B$ and $\partial^2 P/\partial \mu_B^2$, and then demand, at $n_B=2n_0$ and $5n_0$, the interpolating function to match with hadronic and quark equations of state up to the second order derivatives of $P(\mu_B)$. Therefore we have introduced six coefficients, $c_n$'s. In addition the coefficients $c_n$'s must be such that (i) $P(\mu_B)$ is convex and has the positive curvature everywhere; (ii) the growth of $P(\mu_B)$ cannot be too slow for increasing $\mu_B$, otherwise it would violate the causality condition. The pressure curves which do not satisfy these constraints are unphysical. We emphasize that we are using hadronic and quark equations of state {\it only as the boundary conditions}; in particular we do not superpose two different equations of state for the intermediate region, as each equation of state outside of the domain of applicability might show unphysical behaviors.

At this point it is important to emphasize the difference between the three-window modeling and conventional hybrid construction. The latter is based on the assumption that there is a domain where both hadronic and quark matter descriptions are reliable. Then one can describe the first order phase transition from hadron to quark matter by comparing the pressures at given chemical potential, $P_H (\mu_B)$ and $P_Q (\mu_B)$, respectively, and specify the location of the phase transition by the intersection point of two curves. But this procedure by construction strongly biases the possible type of quark equations of state for a given hadronic equations of state, as the chosen $P_Q (\mu_B)$ must intersect $P_H (\mu_B)$ from below. Typical combinations are such that the phase transition happens at $\gtrsim 2n_0$, so in such modeling we implicitly reject some of quark equations of state by the hadronic ones {\it extrapolated beyond the domain of applicability}. Such construction has danger to introduce unphysical constraints on the properties of quark matter.

More conservative way of hybrid construction is the three-window modeling, in which one uses hadronic and quark equations of state only within the domain of applicability, and considers most general curves for intermediate region by using the boundary conditions given by hadronic and quark equations of state. The utility of this consideration is that one no longer has to compare hadronic and quark equations of state directly using the extrapolation of both. This in turn allows us more general class of quark equations of state, some of which have been rejected in the conventional construction. In particular they can accommodate stiff quark equations of state even when we start with soft hadronic equations of state at low density. Indeed unless we demand the quark pressure curves to intersect the hadronic ones from below, there is no strong reason to regard quark equations of state to be soft, as one can already see from the example of a bag model with a small bag constant.

Finally we mention another interpolating procedure in which hadronic equations of state of $\lesssim 1.1 n_0$ and pQCD one of $\gtrsim 50 n_0$ are interpolated \cite{Annala:2017llu}. These works have much more predictive characters than ours including the evaluation of systematic errors; they can be regarded as suppliers for the constraints. In contrast, our modeling starts with the constraints given by several conditions, especially because quark equations of state in neutron star domains cannot be calculated in a quantitative manner from the first principle. Rather our purpose is to delineate the properties of matter satisfying the given constraints. For this reason we will use a schematic quark model with adjustable parameters whose structure is inspired from the hadron and nuclear physics so that one can imagine how the physics at low density is connected to the quark matter region. Also since the model is described in the language of quarks, its connection to the pQCD equations of state is rather easy to imagine. While our use of three-window modeling has more postdictive character, once basic ingredients are specified during the construction of equations of state at zero temperature and at $\beta$-equilibrium, we can predict equations of state in other circumstance, including equations of state at finite temperature, at general isospin fraction, at general strangeness fraction, and so on. Such studies are on-going.

From the next section we discuss the actual construction based on the three-window modeling. The most updated one is the Quark-Hadron-Crossover 2019 (QHC19) \cite{Baym:2019iky} whose tables can be found at [https://compose.obspm.fr/eos/140/].

\section{QHC19: the nuclear part}

For $n_B \lesssim 2n_0$ we use the equation of state constructed by Togashi et al. \cite{Togashi:2017mjp}. The tables are available on the Web at [http://www.np.phys.waseda.ac.jp/EOS/] and the CompOSE archive at [https://compose.obspm.fr/eos/105/].

We briefly summarize the key features of the Togashi equation of state. The way of computations is very similar to the APR which is one of the standard microscopic nuclear equations of state. In short, Togashi's calculations replace some of calculations of the APR with phenomenological ones, but instead significantly extend the range of calculations of the APR to finite temperatures and general proton fractions. As we plan to cover various temperatures, proton fraction, and strangeness fraction in future computations, the Togashi's microscopic computations are suitable for our purposes. In particular since wide domains are computed in the same microscopic forces, there is no matching problem. This is the great advantage, as patching different equations of state with different region of applicability easily violates the thermodynamic consistency. This patching problem becomes more difficult if we have more thermodynamic variables such as temperature, proton fraction, and so on, but this kind of problem is absent in the Togashi equations of state.

The calculations by Togashi et al. are based on a Hamiltonian including two-body interactions extracted by fitting two-nucleon experimental scattering data,  as well as more empirical three-body interactions fit to light nuclei. With these inputs Togashi et al. solved the many-body problem by choosing a variational wave function with parameters determined by minimizing the total energy.  The general form of the wave function is that of a free Fermi gas multiplied by factors that build in two-particle correlations dependent on the relative spin, isospin, and orbital angular momentum of the two-particles. 

The calculations of the Togashi cover from a crust to a nuclear liquid with the same microscopic forces. The crust part is calculated with the Thomas-Fermi method by assuming that a single species of heavy spherical nuclei form a BCC lattice surrounded by a gas of nucleons \cite{Shen:2011,Oyamatsu:1993zz}. The crust extends to density $\simeq 0.625\  n_0$.  As reported in Ref. \cite{Kanzawa:2009}, the properties of neutron star crusts calculated with the Togashi equation of state are consistent with those in previous studies \cite{Baym:1971pw,Negele:1971vb,Douchin:2001sv}. 
 
It is useful to compare the results of Togashi's and the APR at the saturation density. (As for the APR, the numbers cited below are calculated from the fit function in the original APR paper \cite{Akmal:1998cf}.) The saturation energy is 16.1 MeV in Togashi vs. 16.0 MeV in APR, and the incompressibility, 245 vs. 267 MeV in APR. Near $n_0$, SNM in the Togashi equation of state is slightly softer than in APR, as the smaller values of the symmetry energy $J$ = 29.1 vs. 34.0 MeV in APR, and the slope parameter, $L$ = 38.7 vs. 63.2 MeV in APR.
In neutron stars,  the Togashi equation of state is slightly stiffer at higher densities than APR, as seen in the radii, $R_{1.4}$ = 11.6 km for Togashi vs. 11.5 km for APR, and the tidal deformability, $\Lambda_{1.4}$ = 360 vs. 268 (for the crust in APR we need to choose some crust equations of state and we used the Togashi equation of state for $n_{B} \le 0.26n_0$).     

As usual, we couple nuclear equations of state to the charged leptonic one, impose the charge neutrality condition, and then optimize the value of proton fraction to minimize the total energy of the system. 

\section{QHC19: the quark part}

The quark matter part at density less than $\sim 50n_0$ has not been directly accessible from the first principle calculations and one needs some theoretical orientation for a schematic description. Here we use the quark-hadron continuity picture as our guideline. Following this picture the structure of effective models should be imported from those for hadron physics. 

\begin{figure}[tb]
\vspace{0.0cm}
\centering
\includegraphics[scale=0.3]{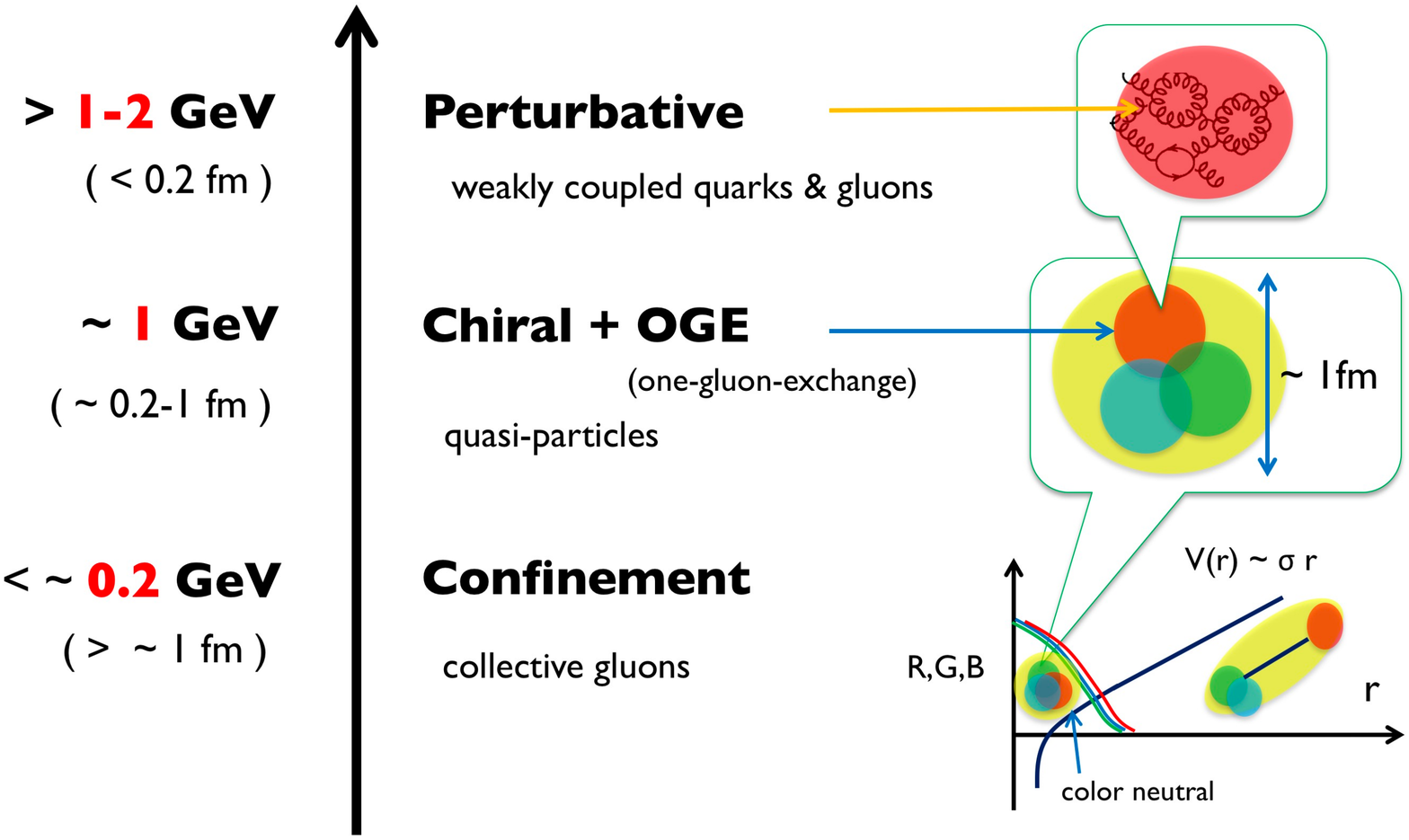}
\caption{
A schematic description of a single hadron.
  }
  \vspace{-0.cm}
   \label{fig:3window_hadrons}
\end{figure}

Before presenting the explicit form of our model, we first summarize the physics supposed to be relevant at $n_B \gtrsim 5n_0$. Here we consider another ``three-window" description of a hadron\footnote{This three-window does not directly match with the three-window used for the description of matter, as the discussions of a hadron do not include the nuclear scale. The nuclear scale is regarded within the confinement scale. } inspired from the arguments originally given by Manohar and Georgi \cite{Manohar:1983md} and later revised by Weinberg \cite{Weinberg:2010bq}. The three window consists of the physics of (i) the confinement for the momentum scale of $\lesssim 0.2$ GeV (or distance $\gtrsim 1$ fm); (ii) the constituent quark dynamics  for 0.2-1 GeV (0.2-1 fm); (iii) the partonic dynamics for $ \gtrsim 1$ GeV ($\lesssim 0.2$ fm). The regime (i) is characterized by the color-electric flux among color charges; for instance the potential between an infinitely heavy quark and an antiquark separated by a distance $r$ is a linear rising one, $V(r) \sim \sigma r$, where $\sigma \sim 1$ GeV/fm is the string tension. In the regime (ii) quarks are treated as if they are quasi-particles with the effective masses of $M_{u,d} \sim 300$ MeV and $M_s \sim 500$ MeV generated through the chiral symmetry breaking, interacting with the other quarks inside of a hadron through few gluon exchanges. This description has been very successful in the hadron spectroscopy in spite of its simplicity. Finally in the regime (iii) we resolve the structure of constituent quarks in space and in time and effective quark masses look the current quark mass $m_{u,d} \sim 5$ MeV and $m_s \sim 100$ MeV. Our modeling below will be based on these pictures.

First of all we have been unable to reliably describe how confining forces change from low to high density. The modeling may be simplified, however, once baryons sufficiently overlap. There the color charge can be almost uniformly distributed so that the length of color-electric flux is short and therefore should not contribute significantly to equations of state (but may contribute to non-bulk quantities involving the transport of colors). A simple model to test this picture is QCD in one-space and one-time dimension in which the color fields are squeezed in one direction and therefore forms a linear rising potential. In the limit of the large number of colors where the string breaking effects are quenched, analyses of this model at finite density \cite{Schon:2000he,Bringoltz:2008iu,Bringoltz:2009ym,Kojo:2011fh} show that the equation of state is largely saturated by the free quark gas contribution, in spite of the presence of confining color-electric flux. With this we restrict the applicability of our modeling to the density beyond $\sim 5n_0$ and omit the detailed discussions about confinement.

Now the most relevant for our studies of the neutron star domain is supposed to be the physics of momentum scale of 0.2-1 GeV (0.2-1 fm). The relevant ingredient is the chiral symmetry breaking and one-gluon-exchange between quarks. We include those effects in a crude way through contact interactions in which momenta in the gluon exchange is fixed to a typical non-perturbative scale of 0.2-1 GeV; as a consequence the description has the cutoff scale of $\sim 1$ GeV, beyond which the coupling constant in front of the contact interactions should drop significantly. Keeping these qualifications in mind, now our effective Hamiltonian is ($\mu_q =\mu_B/3$) \cite{Baym:2017whm}
\begin{eqnarray}
\calH  
= \bar{q} (\rmi \gamma_0 \vec{\gamma}\cdot \vec{\partial} + m -\mu_q \gamma_0)q 
- G_s \sum^8_{i=0} \left[ (\overline{q} \tau_i q)^2 + (\bar{q} \rmi \gamma_5 \tau_i q)^2 \right] 
+ 8 K ( \det\,\!\!_{f} \bar{q}_R q_L + \mbox{h.c.}) \nonumber \\
- H \!\sum_{A,A^\prime = 2,5,7} \!
 \left(\bar{q} \rmi \gamma_5 \tau_A \lambda_{A^\prime} C \bar{q}^T \right) \left(q^T C \rmi \gamma_5 \tau_A \lambda_{A^\prime} q \right) + g_V (\overline{q} \gamma^\mu q)^2
 \,.
 \label{eq:H}
\end{eqnarray}
The first line is the standard Nambu--Jona-Lasinio (NJL) model with $u,d,s$- quarks and responsible for the chiral symmetry breaking. We use the Hatsuda-Kunihiro parameter set \cite{Hatsuda:1994pi} which leads to the dynamically generated quark masses of $M_{u,d} \simeq 336 $ MeV and $M_s \simeq 528$ MeV as functions of the current quark mass matrix $m$, the scalar coupling $G_s$, and the coefficient of the Kobayashi-Maskawa-'t Hooft vertex, $K$. The hadron phenomenology to mesons is successfully described within the Hamiltonian in the first line of Eq.(\ref{eq:H}). Meanwhile the terms in the second line are particularly important when baryons are taken into account. The first term includes the color magnetic interaction for color-flavor-spin antisymmetric S-wave interaction which is attractive. The last term is a phenomenological vector repulsive interaction which is inspired from the $\omega$-meson exchange in nuclear physics. The applications of the NJL model to the physics of dense QCD to the year 2005 are summarized in a review \cite{Buballa:2005}.

While the form of the Hamiltonian is obtained by extrapolating the description of hadron and nuclear physics, in principle the range of parameters $(G_s, K, g_V, H)$ at $n_B \ge 5n_0$ can be considerably different from those used in hadron physics due to e.g. medium screening effects. In strongly correlated region the estimate of medium modifications is difficult; for instance screening masses in two-color QCD, measured in lattice QCD \cite{Hajizadeh:2017ewa}, are qualitatively different from the perturbative behaviors \cite{Kojo:2014vja}. For three-color QCD quantitative estimates on medium modifications are not available, so here we use the neutron star constraints to examine the range of these parameters, and then use them to delineate the properties of QCD matter at $n_B \ge 5n_0$. Below we vary ($g_V, H)$, while assume that $(G_s, K)$ do not change from the vacuum values appreciably; this assumption will be favored posteriori. More elaborated treatment is to explicitly treat the medium running coupling $g_V(\mu_B)$, as demonstrated in Ref.\cite{Fukushima:2015bda}.

Our Hamiltonian for quarks, together with the contributions from leptons, is solved within the mean field approximation. As usual we impose the neutrality conditions for electric and color charges as well as the $\beta$-equilibrium condition. All of these are achieved by introducing the charge and color chemical potentials as Lagrange multipliers. In the mean field treatments we find that the chiral and diquark condensates coexist at $n_B \ge 5n_0$. (For thorough review we refer to \cite{Alford:2007xm}.) For the range of parameters we have explored, the diquark pairing always appears to be the color-flavor-locked (CFL) type at $n_B \ge 5n_0$; other less symmetric pairings such as the 2SC type (in which only up- and down-quark pairs form the condensate) appear only at lower density where the confining effects should be significant.

\section{The roles of effective couplings}

Now we examine the roles of effective interactions by subsequently adding $g_V$ and then $H$ to the standard NJL model \cite{Kojo:2014rca}. First of all, in order to make equations of state stiff, $(G_s, K)_{@5n_0}$ should remain comparable to the size of its vacuum values; the large reduction of these parameters accelerates the chiral restoration that yields contributions similar to the bag constant of 50-200 MeV/fm$^3$, i.e., the positive (negative) contributions to energy (pressure). As a result the significant softening takes place in equations of state. So we fix $(G_s, K)$ to the vacuum values. Actually even after making such choice, the strong 1st order chiral transition takes place at $n_B \sim 2$-$3n_0$ in the standard NJL model, so the equations of state at $n_B\ge 5n_0$ is too soft to pass the $2M_\odot$ constraint. 

\begin{figure}[tb]
\vspace{0.0cm}
\centering
\includegraphics[scale=0.20]{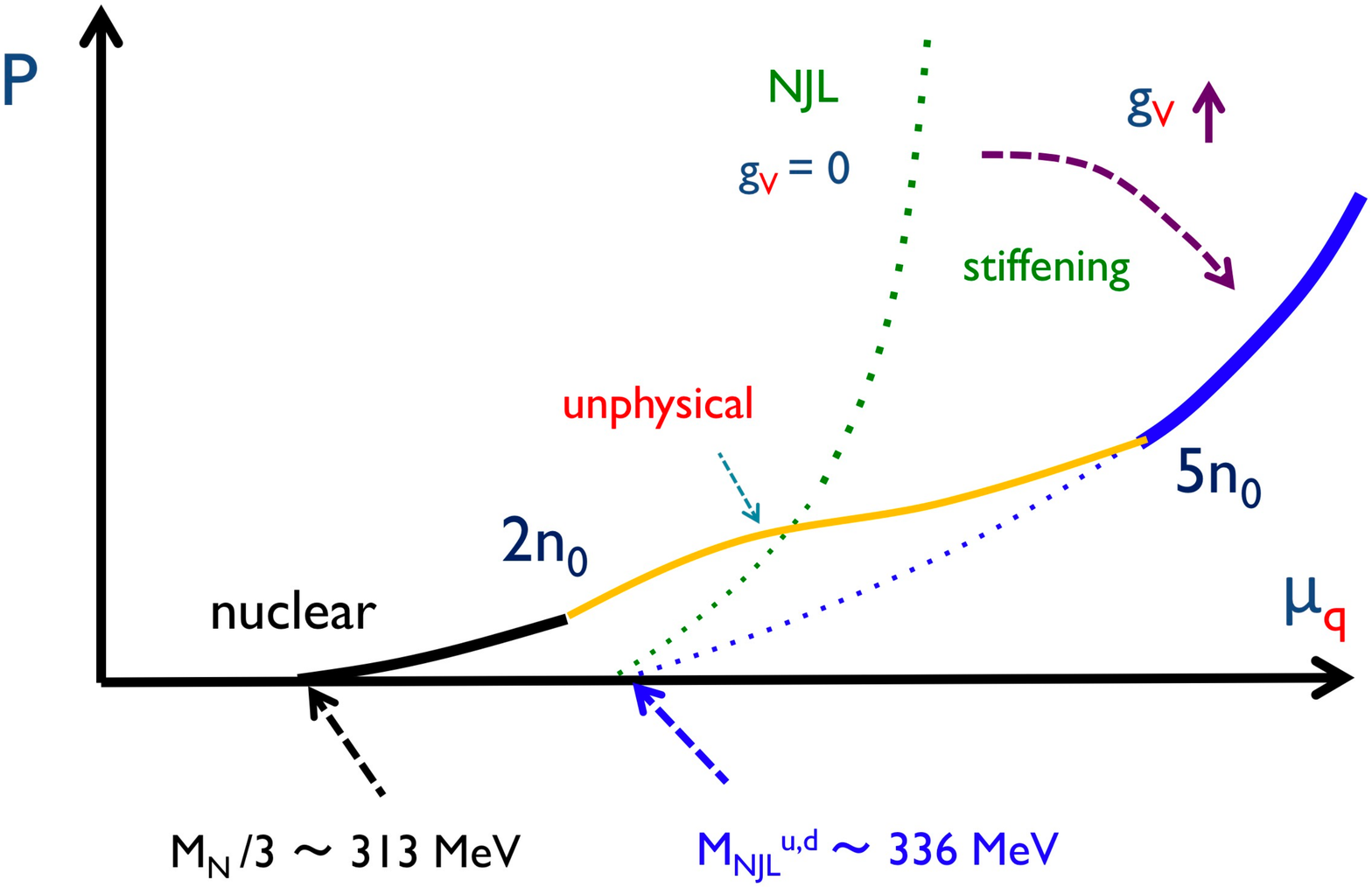}
\hspace{-0.5cm}
\includegraphics[scale=0.20]{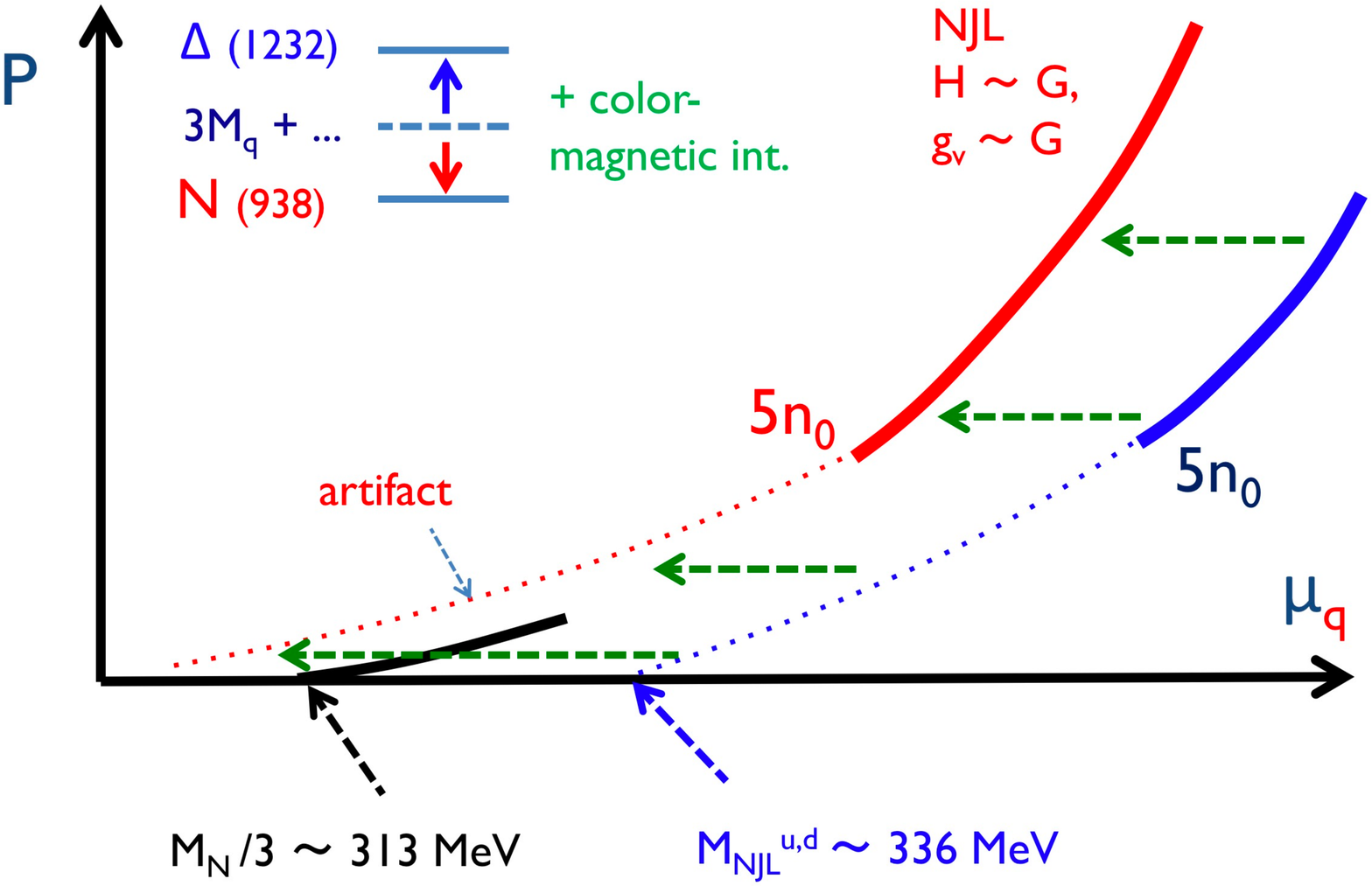}
\hspace{-0.5cm}
\includegraphics[scale=0.20]{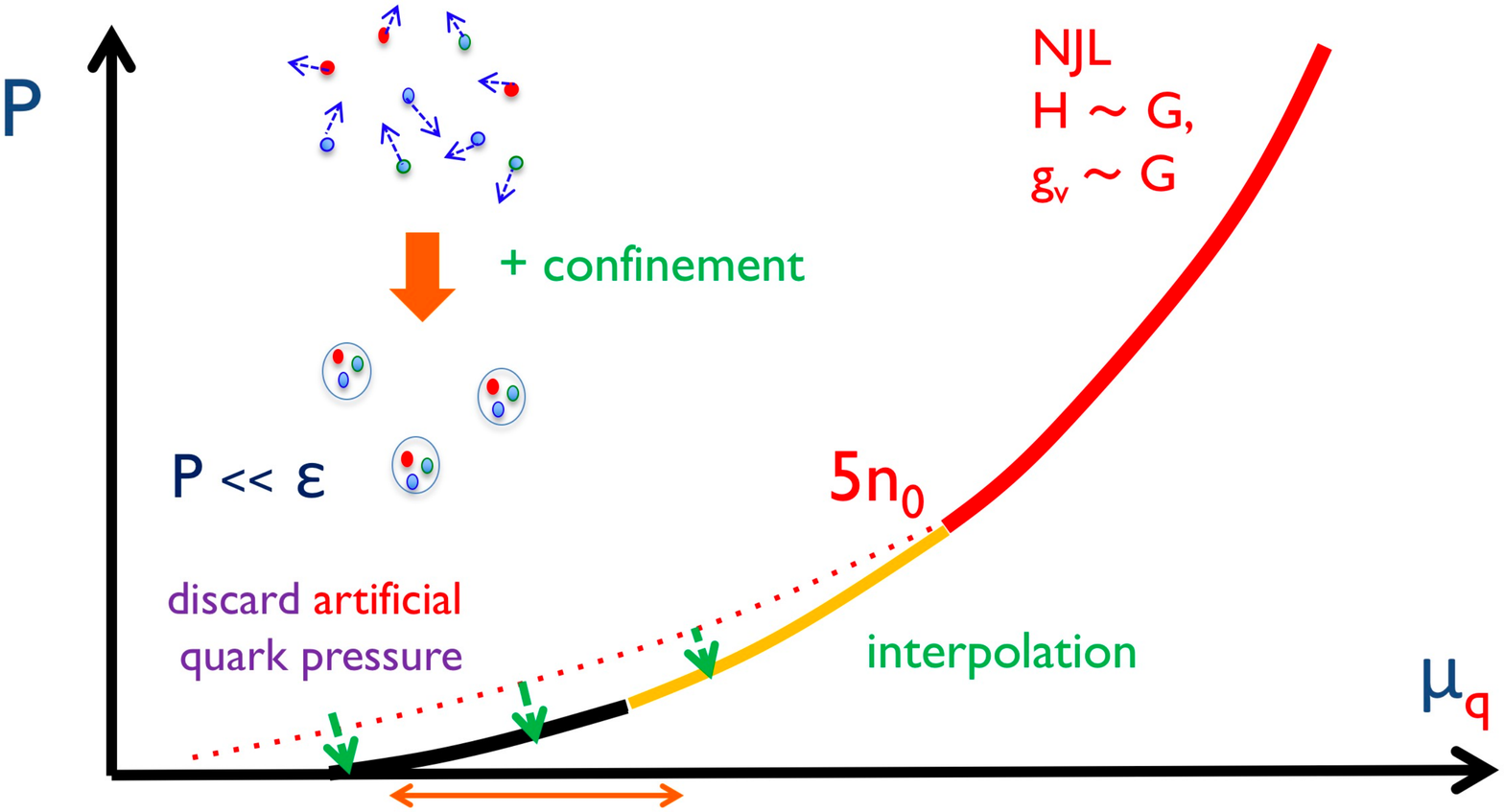}
\caption{The impacts of the vector and color-magnetic interactions as well as of the nuclear constraint. 
  }
  \vspace{-0.5cm}
   \label{fig:gv_H_conf}
\end{figure}

This situation is changed by adding $g_V$. It stiffens the equations of state in two-fold ways. Firstly the repulsive interactions contribute to the stiffening. Secondly, it delays the chiral restoration by tempering the growth of baryon density as a function of $\mu_B$, so that there is no radical softening associated with the chiral restoration. In fact the 1st order transition turns into a crossover in the range of $g_V$ that we explored. The value of $g_V$ large enough to pass the $2M_\odot$ constraint, however, causes another kind of problem in connecting the nuclear and quark model pressure, see the first panel of Fig.\ref{fig:gv_H_conf}; with larger $g_V$ the quark pressure $P(\mu_B)$ tends to appear at higher $\mu_B$ with the less slope, and as a consequence the pressure curve in the interpolation region tends to contain an inflection point at which $\partial^2 P/(\partial \mu_B)^2$ is negative. Such region is thermodynamically unstable so the curve with large $g_V$ must be excluded. Therefore while larger value of $g_V$ is favored to pass the $2M_\odot$ constraint, it generates more mismatch between the nuclear and quark pressure in the $\mu_B$ direction.

Here the color magnetic interactions improve the situation, see the second panel of Fig.\ref{fig:gv_H_conf}. We note that the onset chemical potential of the nuclear pressure is the nucleon mass $\mu_B \simeq 939$ MeV, while for the NJL pressure it is $\mu_B\simeq 3M_{u,d} \simeq 1018$ MeV. In conventional picture of quark models, the nucleon and $\Delta$ masses are split by the color-magnetic interaction, and the nucleon mass is reduced from $3M_{u,d}$. From this viewpoint, the color magnetic interactions naturally induce the overall shift of the NJL pressure toward the lower chemical potential, thus make the matching between the nuclear and quark pressure curves much better.

Finally we comment on the low density behavior after the overall shift. At a given chemical potential the resulting quark matter pressure can be larger than the nuclear one. If we take the quark matter curve for $\lesssim 5n_0$ at its face value, we must reject such parameter sets. As we emphasized, however, we do not regard this situation as a problem at all. In reality the confining effects do not allow dilute quark matter and hence discard artificial pressure. In this picture we even think that the quark matter pressure should approach the nuclear one {\it from above} as we turn on the confining forces.

\section{The range of effective couplings}

Now we explore the range of effective couplings $(g_V, H)$ that fulfils all the required constraints discussed so far.

\begin{figure}[tb]
\vspace{0.0cm}
\centering
\includegraphics[scale=0.7]{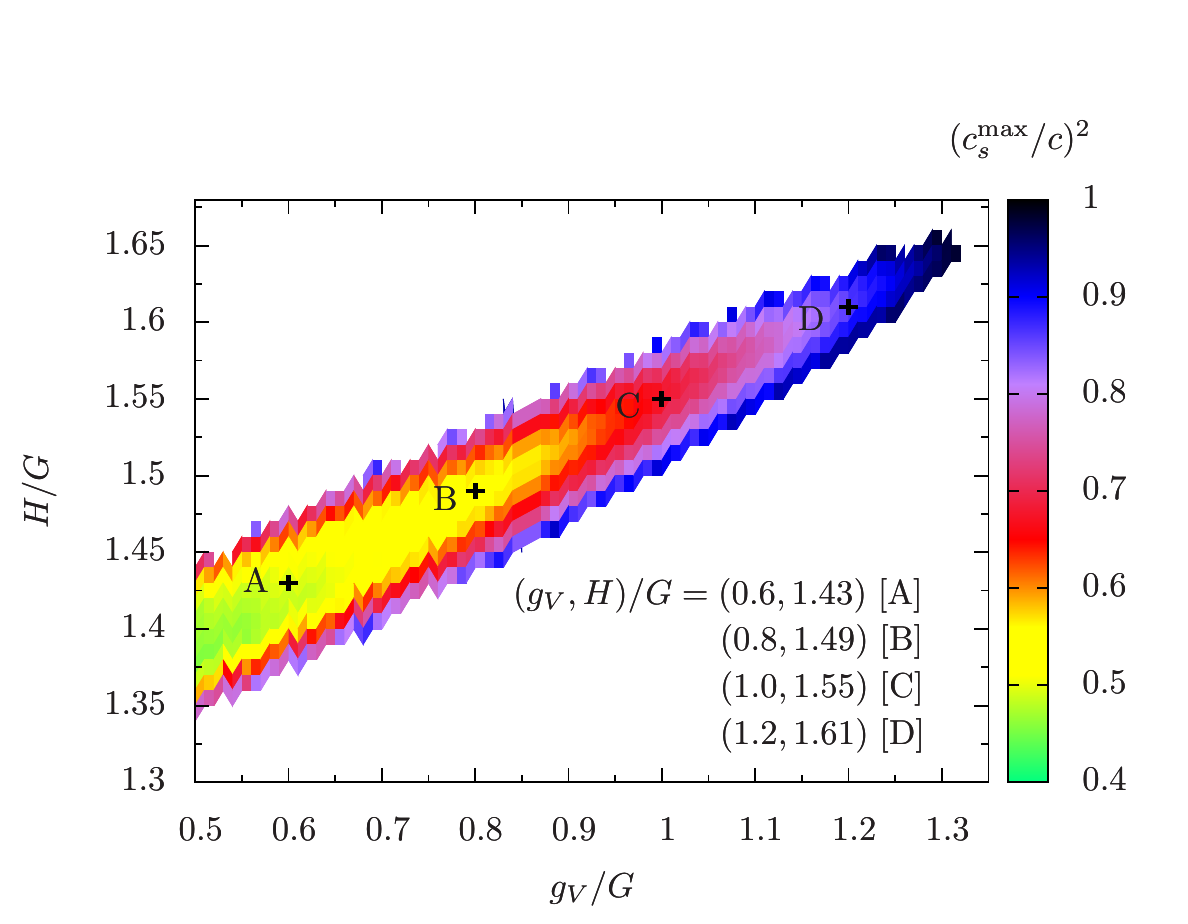}
\hspace{-0.cm}
\includegraphics[scale=0.65]{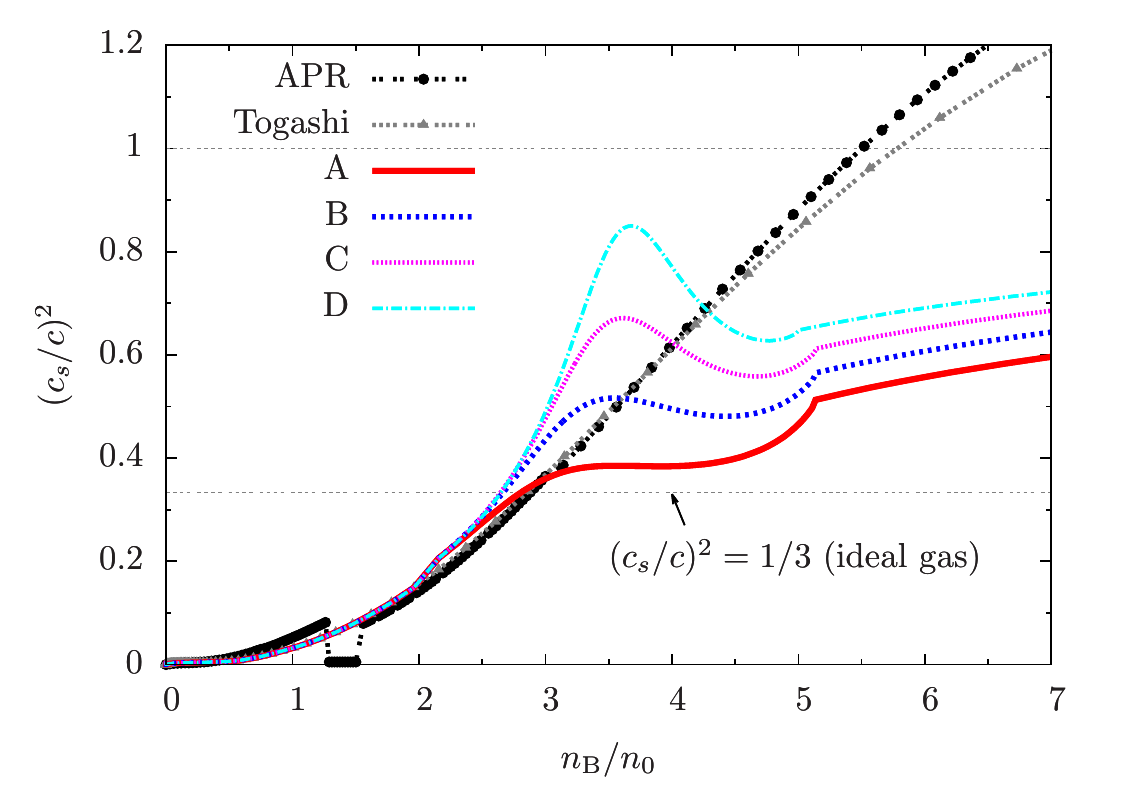}
\caption{ (Left) The maximum value of speed of sound square, $c_s^2$, for a given $(g_V,H)$, found in the interval 2-5$n_0$. The domain left blank is unphysical. The samples A-D are chosen for the right panel which includes also the speed of sound for the APR and Togashi equations of state.
  }
  \vspace{-0.5cm}
   \label{fig:cs2_gv_H}
\end{figure}

First we specify the range where the speed of sound is physical. Shown in the left panel in Fig.\ref{fig:cs2_gv_H} is the maximum of the sound velocity in the interval 2-5$n_0$. In the domain left blank the speed of sound is unphysical. There is a strong correlation between the range of $g_V$ and $H$; for larger $g_V$ we need larger $H$. The reason should be clear from the first two figures in Fig.\ref{fig:gv_H_conf} and the explanations in the main text. As $g_V$ increases the allowed domain for $H$ shrinks and it terminates around $g_V \simeq 1.3\, G_s$. As samples of $(g_V,H)/G_s$, we take (0.6, 1.43) (set A); (0.8, 1.49) (set B); (1.0, 1.55) (set C); and (1.2, 1.61) (set D). The speed of sound square as functions of $n_B$ for sets A-D are shown in the right panel in Fig.\ref{fig:cs2_gv_H}. The results of the APR and Togashi equations of state are also shown. These two nuclear equations of state violate the causality for $\gtrsim 5n_0$. Meanwhile the QHC19 for the sets A-D have peaks or bumps for 2-5$n_0$. The slowly growing behavior beyond 5$n_0$ are likely artifacts of keeping the form of contact interactions. We expect that in more realistic treatment of gluon exchanges the contact interactions are gradually weakened as the density increases, and the speed of sound square approaches 1/3. This part is being investigated \cite{Song:2019qoh}. The Schwinger-Dyson approach can be used to improve the connection to the pQCD equations of state \cite{Bai:2017wvk}.

\begin{figure}[tb]
\vspace{0.0cm}
\centering
\includegraphics[scale=0.65]{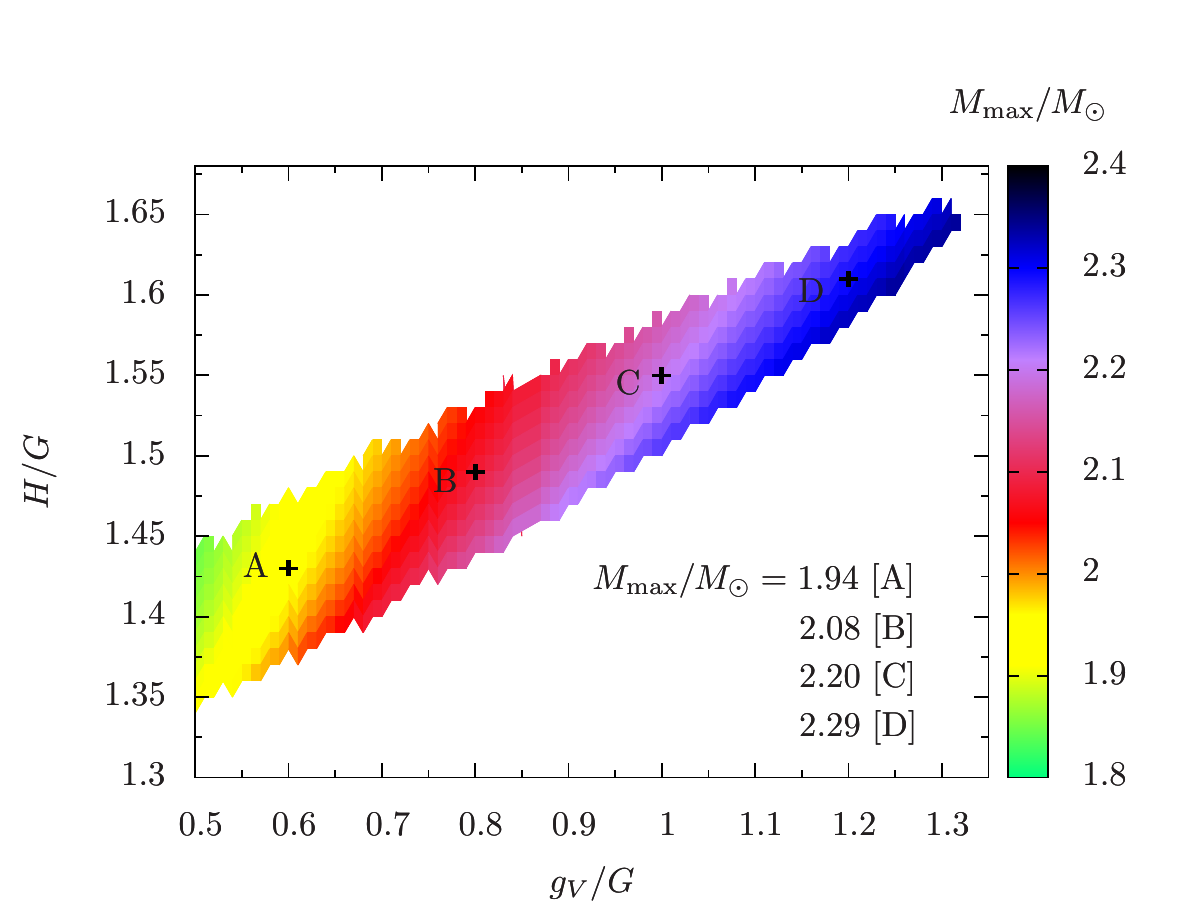}
\hspace{-0.cm}
\includegraphics[scale=0.65]{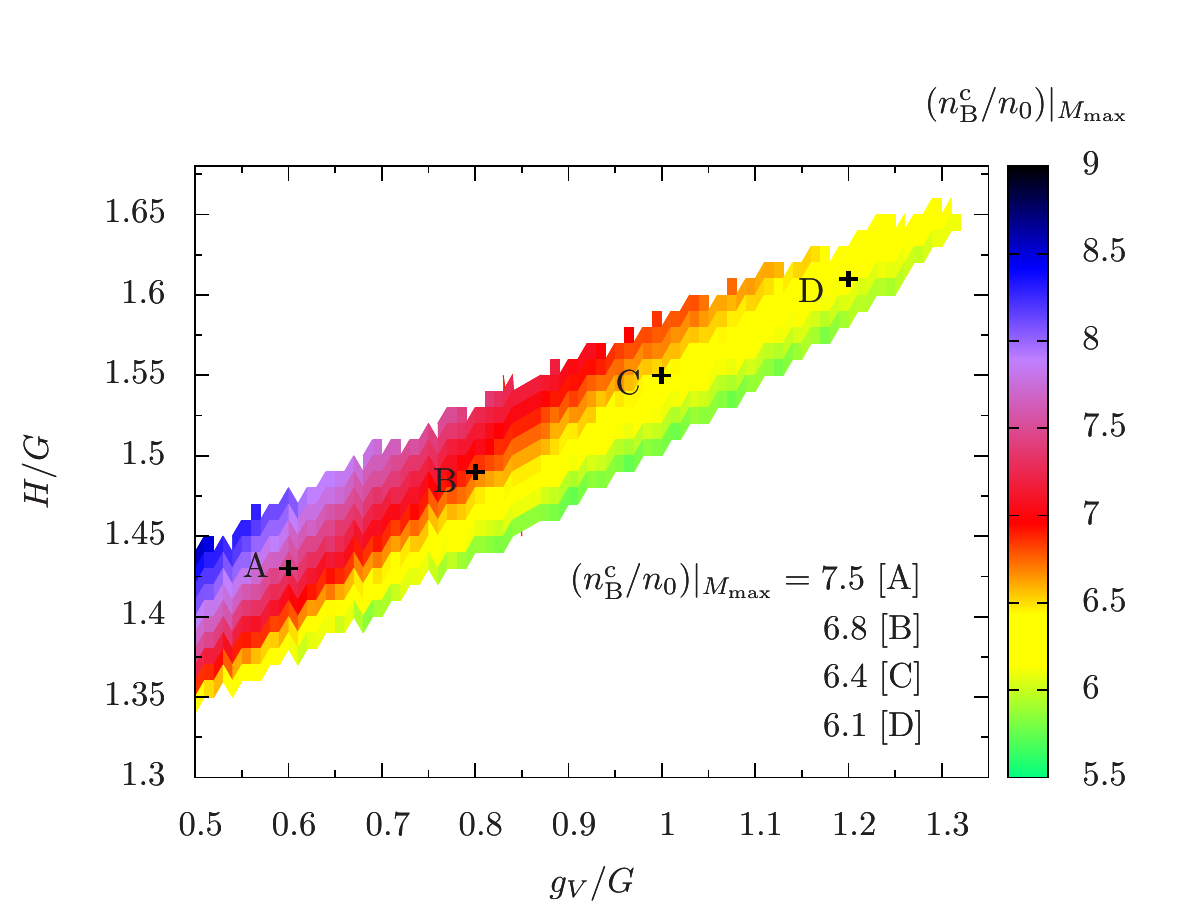}
\caption{ The maximum mass (left) and the core baryon density (right) of neutron stars for a given $(g_V,H)$. 
  }
  \vspace{-0.5cm}
   \label{fig:Mmax}
\end{figure}

Next we examine the maximum of the neutron star masses, as shown in the left panel of Fig.\ref{fig:Mmax}. It shows that $g_V/G_s$ should be larger than 0.6-0.7, otherwise the maximum mass does not exceed $2M_\odot$. Correspondingly $H/G_s$ must be bigger than $\sim 1.35$. We found that such choice of $H$ leads to the diquark gap of $\sim$ 200-250 MeV at $n_B \sim 5n_0$. Within our approach the largest maximum mass is found to be $\simeq 2.35M_\odot$ where the domain of $H$ is about to terminate. 

Shown in the right panel of Fig.\ref{fig:Mmax} is the core baryon density, $n_B^c$. For a given $g_V$, a larger $H$ leads to the larger $n_B$. This is natural because the diquark pairing favors the larger Fermi surface to create pairs as much as possible for the reduction of the total energy of the system. For the similar reasons, larger $g_V$ for a given $H$ reduces $n_B^c$. It is remarkable that in the entire domain we have explored the core density reaches the baryon density larger than $5n_0$. Meanwhile with the $2M_\odot$ constraint the core density $n_B^c$ does not exceed $\simeq 8n_0$. But it should be kept in mind that, if we use nuclear equations of state stiffer than the Togashi's around 1-2$n_0$, the core density can be smaller than the current estimate as the stiffness before reaching $5n_0$ makes the neutron star bigger in size.

\begin{figure}[tb]
\vspace{0.0cm}
\centering
\includegraphics[scale=0.78]{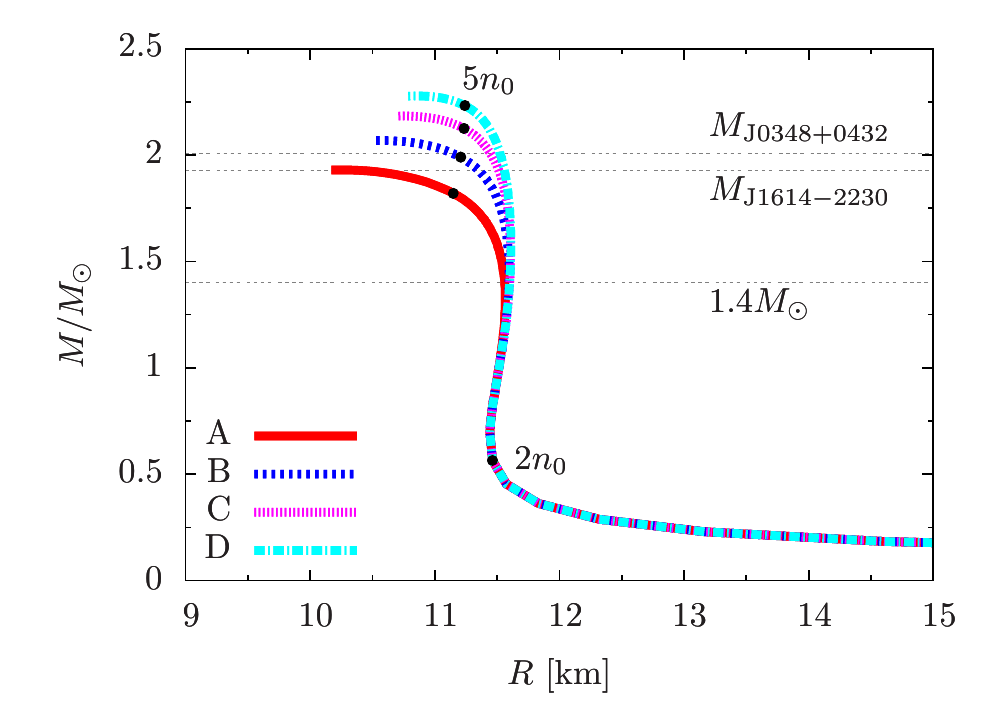}
\hspace{-0.cm}
\includegraphics[scale=0.6]{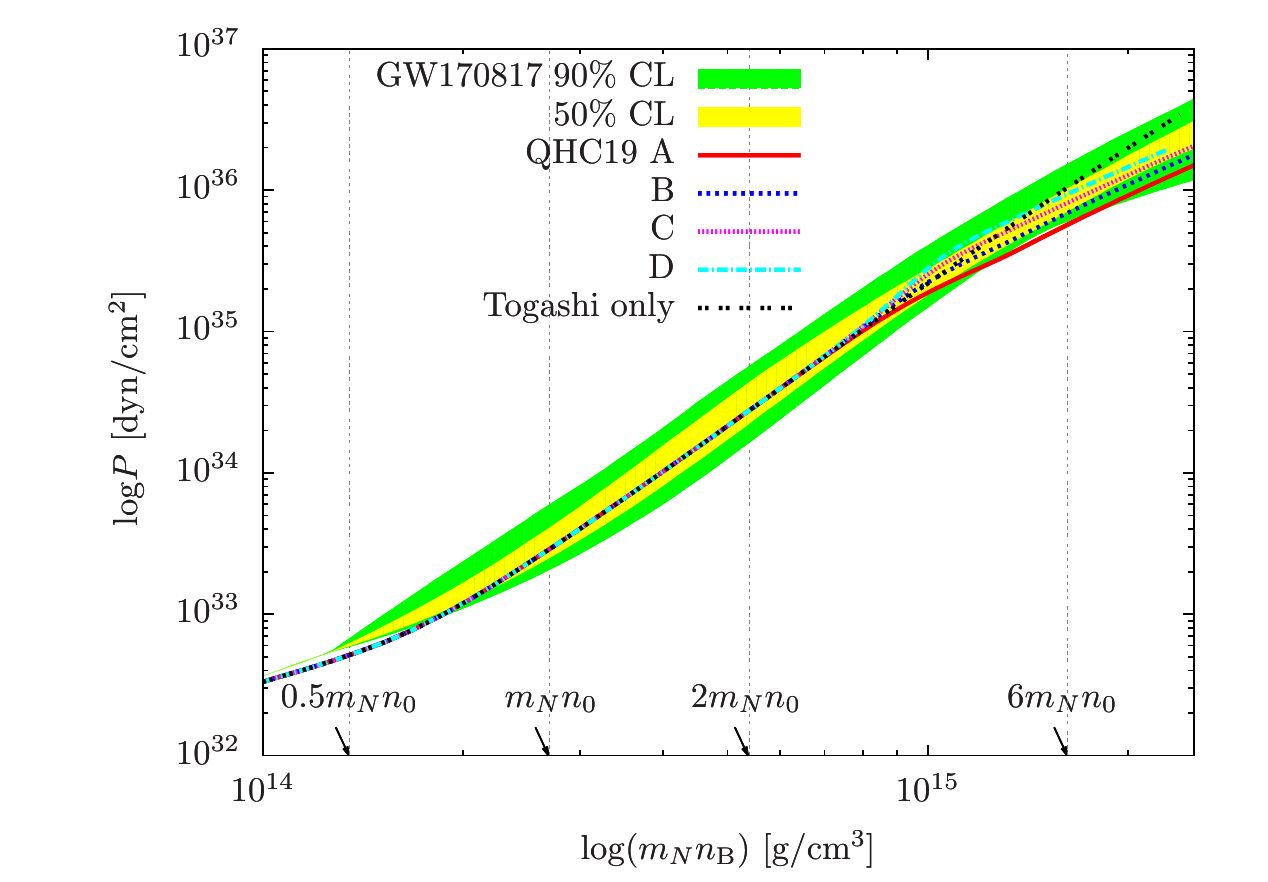}
\caption{ (Left) The mass-radius relations for the sets A-D. (Right) The equations of state inferred from the bayesian analysis by aLIGO collaboration. They are compared to QHC19 for the set A-D. 
  }
  \vspace{-0.5cm}
   \label{fig:M-R_LIGO}
\end{figure}

The $M$-$R$ relations are shown in Fig.\ref{fig:M-R_LIGO} for the sets A-D, together with points indicating where the core densities of $2n_0$ and $5n_0$ are reached. The radius $R_{1.4}$ of a neutron star is $\simeq 11.6$ km, mainly determined by the Togashi equations of state. The quark model parameters varied for the entire allowed range affect $R_{1.4}$ by $\simeq 0.5$ km at most. Meanwhile the maximum mass is sensitive to the quark model parameters, as it should.

Shown in Fig.\ref{fig:M-R_LIGO} is a band for the equations of state inferred from the bayesian analysis by aLIGO collaboration \cite{Ligo-ns:2018a,Ligo-ns:2018b}. We put the equations of state QHC19 for the set A-D. The analyses by the aLIGO took the SLy equations of state at low density, and use the constraints from the tidal deformability and the $2M_\odot$. By construction the QHC19 is largely consistent with the band. But more important is how the QHC19 has been made from our quark model as explained in the previous section. We have clarified the relation between the equation of state and effective interactions which are familiar for people working on the QCD phase diagram and hadron physics. Although the estimate of these interactions are still rather crude due to our simplifications in treating the gluon exchange, the overall conclusion is that the strength of effective interactions is as strong as those in vacuum.

\section{Discussion: strangeness and speed of sound}

We close this article with several remarks for the future improvements of our modeling.

(i) As we consider the interpolation only for the function $P(\mu_B)$, our method does not directly specify the particle contents in the domain 2-5$n_0$. But still we can infer what would happen by looking at the flavor content at $\gtrsim 5n_0$. We found that the QHC19 with $(g_V, H)$ compatible with the constraints have (almost) equal numbers of up-, down-, and strange-quarks at $\ge 5n_0$. This is in part due to the fact that the baryon chemical potential $\mu_B$ for $\gtrsim 5n_0$ already exceeds 1.5 GeV ($\simeq 3M^{vac}_s$), and beyond the threshold the density of strangeness can quickly catch up the densities of up- and down-quarks. In addition the diquark pair correlations are the strongest in the CFL form in which the diquark pairing minimizes the representations of flavor, color, and spin in the system to reduce the corresponding charges. The equal population of $u,d,s$-quarks makes the wavefunction of the system the $SU(3)$ flavor singlet. Since the electric charges of up-, down-, and strange-quarks are $2/3$, $-1/3$, and $-1/3$, respectively, the physics of QCD by itself achieves the charge neutrality condition and hence there is almost no charged leptons. This is in sharp contrast to the nuclear matter description based on particle contents, $npe\mu$, extrapolated to $\sim 5n_0$. In order to connect the nuclear and quark matter descriptions we must assume the lepton fraction to decrease toward the quark matter domain. The onset of the strangeness is important for the neutron star cooling as it is sensitive to the lepton fractions, the presence of gaps, and so on. The strangeness is also important to examine the relationship between the neutron star physics and the physics of heavy ion collisions at low energy; in the former the time scale is sufficiently long for up- and down-quarks to produce the strangeness via weak processes, while in the latter the time scale is too short for the weak processes and the net strangeness is fixed to $\sim 0$ as in the initial condition before the collisions.

(ii) We have not attempted to directly model how baryonic matter transforms into quark matter. This is the most important and difficult question. There are a number of researches toward a unified description from the structure of hadrons to quark matter matter properties \cite{Zhu:2018vwn,Ma:2018qkg,McLerran:2018hbz}. Taking a unified description there should be no matching problem between different equations of state although the actual calculations become complex and some assumptions must be introduced. The Refs.\cite{Ma:2018qkg,McLerran:2018hbz} particularly focus on the question of how the supposed enhancement of the sound velocity for the nuclear-to-quark matter region can be explained. In \cite{Ma:2018qkg} the baryons and baryonic matter are based on Skyrmions which reveal the structural changes around 2-3$n_0$. Meanwhile the description of \cite{McLerran:2018hbz} is based on the quarkyonic matter hypothesis \cite{McLerran:2007qj,Hidaka:2008yy,Kojo:2009ha} in which the matter has the quark Fermi sea with the baryonic Fermi surface which is due to the confining forces. The latter scenario is more directly related to our descriptions as it addresses how the quark or nucleon Fermi seas are formed. The idea is fundamental and we briefly sketch it here (although it will be presented in a different way than the original).

At low baryon density $n_B \ll \Lambda^3_{QCD}$, quarks are trapped in baryons and the quarks have typical momenta of $k_q \sim \Lambda_{QCD}$. But momenta of $N_c$-quarks are added up to much smaller total momentum of a baryon, $K_B$.
Thus 
\beq
\left\langle \sum_{i=1}^{N_c} \vec{k}_q^{(i)} \right\rangle  \sim \vec{K}_B \ll \Lambda_{QCD} \,,~~~~~~~ \left\langle \left(\vec{k}_q^{(i)} - \langle \vec{k}_q^{(i)} \rangle \right)^2 \right\rangle \sim \Lambda_{QCD}^2 \,.
\eeq
Quarks do not directly contribute to the pressure as the confinement forbids, but do contribute to the energy of baryons. Hence in this regime $P \sim n_B^{5/3}/(N_c \Lambda_{QCD} )\ll \varepsilon \sim (N_c \Lambda_{QCD} ) n_B$ and the equation of state is very soft. Adding baryons to the system, the baryon Fermi sea is filled from low momentum states with the occupation probability $f_B (K_B \lesssim \Lambda_{QCD}) \sim 1$, while the quark Fermi sea at low momenta is not fully occupied as quarks inside of a baryon have rather broad momentum distribution. Hence the Pauli blocking at the quark level is not important here.

As the baryon density approaches $\sim \Lambda^3_{QCD}$, however, the quark momentum distributions of baryons are accumulated for low momentum states and start to saturate the possible occupation number. Here $K_B \sim k_q \sim \Lambda_{QCD}$. At this stage one should address not only the Pauli blocking at the level of baryon quantum numbers but also the internal quantum numbers; baryons cannot be added to the system as in free fermions (even if we artificially switch off the baryon-baryon interactions).

After quarks saturate the low momentum states, the quark Fermi surface must be pushed outward and quarks start to have momentum $k_q \gtrsim \Lambda_{QCD}$. Meanwhile if $N_c$-quarks move collectively in the same direction as a baryon then the total momentum is $K_B \sim N_c k_q$. But the growth of the baryon number must be consistent in the quark and baryon descriptions. This demands the baryon occupation number $f_B(K_B)$ at $K_B \gtrsim \Lambda_{QCD}$ to satisfy
\beq
n_B ~\sim~ f_B(K_B) K_B^3 ~ \sim~ k_q^3 ~~~\rightarrow~~~ f_B(K_B)  ~\sim~ N_c^{-3} \,.
\eeq
With this occupation number probability, the baryonic description keeps the baryon number of $O(1)$ while induce larger energy density as baryons occupy high momentum states. In quasi-particle picture for baryons it is
\beq
\varepsilon ~\sim~ \int^{N_c k_q} d{\vec{k}} ~f_B(K_B) \sqrt{K_B^2 +M_B^2 } ~\sim~ N_c^{-3} (N_c k_q)^4 ~\sim~ N_c n_B^{4/3} \,,
\eeq
which readily leads to $P\sim N_c n_B^{4/3}$.
This estimate is consistent with that based on quark calculations; we could reproduce it {\it within the baryonic picture} by assigning the correct occupation probability of baryonic states. The large $N_c$ arguments are used to characterize the overall tendency of the occupation probability, leaving aside the other unimportant factors.

Thus, around the density where the quark occupation probability starts to get saturated, within small window in $n_B$ the pressure changes radically as $\sim n^{5/3}/(N_c \Lambda_{QCD} )$ to $\sim N_c n_B^{4/3}$ while the energy density changes much more smoothly. In intuitive terms: whether or not the matter is baryonic or quark matter, the system always observes the quark contributions to the energy density; at low density the quark energy is observed in the form of the baryon mass while at high density its contribution is directly observed in terms of kinetic energy, and hence the radical change {\it in energy} does not take place from low to high densities. The situation is totally different in {\it pressure}; the system does not observe the quark contributions to the pressure because of the confinement until the density becomes high enough. At some point the system starts to observe the large quark pressure which have been hidden inside of baryons. This occurs within small window of $n_B$ or $\varepsilon$, and therefore leads to a large $dP/d\varepsilon$ in this transition regime. After the transition is completed both the pressure and energy change smoothly, and hence $dP/d\varepsilon$ is relaxed back to the quark gas limit.

Actually having a peak in $c_s^2$ exceeding $1/3$ is not the sufficient condition to pass the $2M_\odot$ constraint; this can be recognized from Figs.\ref{fig:cs2_gv_H} and \ref{fig:Mmax}, where some set of parameters lead to the peak in speed of sound close to the causal limit but does not lead to the maximum mass exceeding $2M_\odot$. But the above arguments do help us to understand why the microscopic considerations based on quarks can lead to the peak in the speed of sound without invoking the nuclear interactions comparable to the nucleon mass. We also expect the quark descriptions to solve the problems found in pure hadronic descriptions, including the hyperon softening, but this will be discussed elsewhere.

To summarize, the physics of neutron stars, heavy ion collisions, and nuclear experiments at low energy are very helpful for the understanding of the structure of cold, dense QCD.

\section{Acknowledgement}

The author thanks the organizers for this enjoyable meeting. He also acknowledges his collaborators, G. Baym, K. Fukushima, S. Furusawa, T. Hatsuda, P. Powell, Y. Song, T. Takatsuka, and H. Togashi, for helpful discussions on this topic. The author also thanks L. McLerran for his explanations about the peak of the sound velocity. The author was supported in part by NSFC grant 11650110435 and 11875144, and also by a grant from the Simons Foundation with which in 2018 he could perform some works at Aspen Center for Physics, which is supported by National Science Foundation grant PHY-1607611. He acknowledges the KMI in the Nagoya University for the visiting program during which this article was finalized.



\begin{thebibliography}{1}

\bibitem{fonseca} E. Fonseca et al.,  
Astrophys.\ J.\  {\bf 832}, 167:1-22 (2016).

\bibitem{Demorest} P. Demorest, T. Pennucci, S. M. Ransom, R. M. S. E., and J. W. T. Hessels, 
Nature (London) \textbf{467}, 1081-3 (2010).

\bibitem{Antoniadis2013}J. Antoniadis, et al., 
Science \textbf{340}, 1233232 (2013).

\bibitem{2.17mass} H. T. Cromartie et al., arXiv:1904.06759 [astro-ph.HE].

\bibitem{timing} A. L. Watts, N. Andersson, D. Chakrabarty, M. Feroci, J. Hebeler, G.
Israel, F. K. Lamb, M. C. Miller, S. Morsink, F. \"Ozel, A.
Patruno, J. Poutanen, D. Psaltis, A. Schwenk, A. W. Steiner, L. Stella,
L. Tolos, and M. van der Klis, 
Rev. Mod. Phys. {\bf 88}, 021001: 1-23 (2016).



\bibitem{TheLIGOScientific:2017qsa}
  B.~P.~Abbott {\it et al.} [LIGO Scientific and Virgo Collaborations], 
  Phys.\ Rev.\ Lett.\  {\bf 119} (2017) no.16,  161101.

\bibitem{GW170817A} B. P. Abbott et al., 
Ap. J. Letters {\bf 848}, L12:1-59 (2017), and following papers in Ap. J. Letters {\bf 848}.

\bibitem{Monitor:2017mdv}
  B.~P.~Abbott {\it et al.} [LIGO Scientific and Virgo and Fermi-GBM and INTEGRAL Collaborations],
  Astrophys.\ J.\  {\bf 848} (2017) no.2,  L13.

\bibitem{Coulter:2017wya}
  D.~A.~Coulter {\it et al.},
  Science {\bf 358} (2017) 1556.
  
  



\bibitem{Lattimer:2012xj}
  J.~M.~Lattimer and Y.~Lim,
  Astrophys.\ J.\  {\bf 771} (2013) 51.

\bibitem{Oertel:2016bki}
  M.~Oertel, M.~Hempel, T.~Klähn and S.~Typel,
  Rev.\ Mod.\ Phys.\  {\bf 89} (2017) no.1,  015007.



\bibitem{Li:2008gp}
  B.~A.~Li, L.~W.~Chen and C.~M.~Ko,
  Phys.\ Rept.\  {\bf 464} (2008) 113.

\bibitem{Danielewicz:2002pu}
  P.~Danielewicz, R.~Lacey and W.~G.~Lynch,
  Science {\bf 298} (2002) 1592.



\bibitem{Russotto:2011hq}
  P.~Russotto {\it et al.},
  Phys.\ Lett.\ B {\bf 697} (2011) 471.
  
  \bibitem{Russotto:2016}
  P.~Russotto {\it et al.},
  Phys.\ Rev.\ D {\bf 79} (2009) 014004.





\bibitem{Baym:2017whm}
  G.~Baym, T.~Hatsuda, T.~Kojo, P.~D.~Powell, Y.~Song and T.~Takatsuka,
  Rept.\ Prog.\ Phys.\  {\bf 81} (2018) no.5,  056902.

\bibitem{Baym:2019iky}
  G.~Baym, S.~Furusawa, T.~Hatsuda, T.~Kojo and H.~Togashi,
  arXiv:1903.08963 [astro-ph.HE].
  
  

\bibitem{Akmal:1998cf}
  A.~Akmal, V.~R.~Pandharipande and D.~G.~Ravenhall,
  Phys.\ Rev.\ C {\bf 58} (1998) 1804.
  
\bibitem{Togashi:2017mjp}
  H.~Togashi, K.~Nakazato, Y.~Takehara, S.~Yamamuro, H.~Suzuki and M.~Takano,
  Nucl.\ Phys.\ A {\bf 961} (2017) 78.



\bibitem{Kurkela:2009gj}
  A.~Kurkela, P.~Romatschke and A.~Vuorinen,
  Phys.\ Rev.\ D {\bf 81} (2010) 105021.
\bibitem{Fraga:2001id}
  E.~S.~Fraga, R.~D.~Pisarski and J.~Schaffner-Bielich,
  Phys.\ Rev.\ D {\bf 63} (2001) 121702.
\bibitem{Freedman:1976ub}
  B.~A.~Freedman and L.~D.~McLerran,
  Phys.\ Rev.\ D {\bf 16} (1977) 1169.




\bibitem{Hands:2010gd}
  S.~Hands, S.~Kim and J.~I.~Skullerud,
  Phys.\ Rev.\ D {\bf 81} (2010) 091502.
\bibitem{Cotter:2012mb}
  S.~Cotter, P.~Giudice, S.~Hands and J.~I.~Skullerud,
  Phys.\ Rev.\ D {\bf 87} (2013) no.3,  034507.

\bibitem{Braguta:2016cpw}
  V.~V.~Braguta, E.-M.~Ilgenfritz, A.~Y.~Kotov, A.~V.~Molochkov and A.~A.~Nikolaev,
  Phys.\ Rev.\ D {\bf 94} (2016) no.11,  114510.

\bibitem{Hajizadeh:2017ewa} 
  O.~Hajizadeh, T.~Boz, A.~Maas and J.~I.~Skullerud,
  EPJ Web Conf.\  {\bf 175}, 07012 (2018).

\bibitem{Boz:2018crd}
  T.~Boz, O.~Hajizadeh, A.~Maas and J.~I.~Skullerud,
  arXiv:1812.08517 [hep-lat].

\bibitem{Masuda:2012kf}
  K.~Masuda, T.~Hatsuda and T.~Takatsuka,
  Astrophys.\ J.\  {\bf 764} (2013) 12.
\bibitem{Masuda:2012ed}
  K.~Masuda, T.~Hatsuda and T.~Takatsuka,
  PTEP {\bf 2013} (2013) no.7,  073D01.

\bibitem{Kojo:2014rca}
  T.~Kojo, P.~D.~Powell, Y.~Song and G.~Baym,
  Phys.\ Rev.\ D {\bf 91} (2015) no.4,  045003.

\bibitem{Kojo:2015fua}
  T.~Kojo,
  Eur.\ Phys.\ J.\ A {\bf 52} (2016) no.3,  51.

\bibitem{Demorest:2010bx}
  P.~Demorest, T.~Pennucci, S.~Ransom, M.~Roberts and J.~Hessels,
  Nature {\bf 467} (2010) 1081.
  
\bibitem{Antoniadis:2013pzd}
  J.~Antoniadis {\it et al.},
  Science {\bf 340} (2013) 6131.

\bibitem{Ozel:2016oaf}
  F.~Ozel and P.~Freire,
  Ann.\ Rev.\ Astron.\ Astrophys.\  {\bf 54} (2016) 401.

\bibitem{Lindblom1992} L. Lindblom, 
Ap. J. {\bf 398},  569-573 (1992). 

\bibitem{Fujimoto:2017cdo}
  Y.~Fujimoto, K.~Fukushima and K.~Murase,
  Phys.\ Rev.\ D {\bf 98} (2018) no.2,  023019.

\bibitem{Fujimoto:2019hxv}
  Y.~Fujimoto, K.~Fukushima and K.~Murase,
  arXiv:1903.03400 [nucl-th].


\bibitem{Witten:1984rs}
  E.~Witten,
  Phys.\ Rev.\ D {\bf 30} (1984) 272.


\bibitem{Lattimer:2006xb}
  J.~M.~Lattimer and M.~Prakash,
  Phys.\ Rept.\  {\bf 442} (2007) 109.



\bibitem{Maslov:2018ghi}
  K.~Maslov, N.~Yasutake, A.~Ayriyan, D.~Blaschke, H.~Grigorian, T.~Maruyama, T.~Tatsumi and D.~N.~Voskresensky,
  arXiv:1812.11889 [nucl-th].

\bibitem{Benic:2014jia}
  S.~Benic, D.~Blaschke, D.~E.~Alvarez-Castillo, T.~Fischer and S.~Typel,
  Astron.\ Astrophys.\  {\bf 577} (2015) A40.

\bibitem{Most:2018eaw} 
  E.~R.~Most, L.~J.~Papenfort, V.~Dexheimer, M.~Hanauske, S.~Schramm, H.~Stöcker and L.~Rezzolla,
  Phys.\ Rev.\ Lett.\  {\bf 122}, no. 6, 061101 (2019).

\bibitem{Bauswein:2018bma} 
  A.~Bauswein, N.~U.~F.~Bastian, D.~B.~Blaschke, K.~Chatziioannou, J.~A.~Clark, T.~Fischer and M.~Oertel,
  Phys.\ Rev.\ Lett.\  {\bf 122}, no. 6, 061102 (2019).

\bibitem{Han:2018mtj} 
  S.~Han and A.~W.~Steiner,
  Phys.\ Rev.\ D {\bf 99}, no. 8, 083014 (2019).

\bibitem{Fischer:2017lag} 
  T.~Fischer {\it et al.},
  Nat.\ Astron.\  {\bf 2}, no. 12, 980 (2018).


\bibitem{nicer}  K. C. Gendreau et al.,  ``The Neutron star Interior Composition Explorer (NICER): design and development,"
Proc. SPIE 9905, Space Telescopes and Instrumentation 2016: Ultraviolet to Gamma Ray, 99051H (July 22, 2016).

\bibitem{michi}  M. Baub\"ock, D. Psaltis, and F. \"Ozel,  
Astrophys. J. {\bf 811}, 144:1-8 (2015).

\bibitem{Miller:2016kae}
  M.~C.~Miller,  
  Astrophys.\ J.\  {\bf 822}, 27:1-7 (2016).

\bibitem{ozel-nicer} F. \"Ozel, D. Psaltis,, Z. Arzoumanian, S. Morsink, and M. Baub\"ock, 
Astrophys. J. {\bf 832}, 92:1-8 (2016).




\bibitem{Foley:2019evo}
  R.~J.~Foley {\it et al.},
  arXiv:1903.04553 [astro-ph.HE].

\bibitem{Hinderer:2015} Hinderer, T. 2008, Astrophys. J. 677 1216-20; erratum 2009, Astrophys. J. 697, 964


\bibitem{Margalit:2017dij}
  B.~Margalit and B.~D.~Metzger,
  Astrophys.\ J.\  {\bf 850} (2017) no.2,  L19.

\bibitem{Rezzolla:2017aly}
  L.~Rezzolla, E.~R.~Most and L.~R.~Weih,
   Astrophys.\ J.\ Lett.\  {\bf 852} (2018) L25.

\bibitem{Ruiz:2017due}
  M.~Ruiz, S.~L.~Shapiro and A.~Tsokaros,
  Phys.\ Rev.\ D {\bf 97} (2018) no.2,  021501.

\bibitem{Shibata:2017xdx}
  M.~Shibata, S.~Fujibayashi, K.~Hotokezaka, K.~Kiuchi, K.~Kyutoku, Y.~Sekiguchi and M.~Tanaka,
  Phys.\ Rev.\ D {\bf 96} (2017) no.12,  123012.

\bibitem{Yu:2017syg}
  Y.~W.~Yu, L.~D.~Liu and Z.~G.~Dai,
  Astrophys.\ J.\  {\bf 861} (2018) no.2,  114.

\bibitem{Bauswein:2017vtn} 
  A.~Bauswein, O.~Just, H.~T.~Janka and N.~Stergioulas,
  Astrophys.\ J.\  {\bf 850}, no. 2, L34 (2017).

\bibitem{Radice:2017lry}
  D.~Radice, A.~Perego, F.~Zappa and S.~Bernuzzi,
  Astrophys.\ J.\  {\bf 852} (2018) no.2,  L29.


\bibitem{Annala:2017llu}
  E.~Annala, T.~Gorda, A.~Kurkela and A.~Vuorinen,
  Phys.\ Rev.\ Lett.\  {\bf 120} (2018) no.17,  172703.
  

\bibitem{Alford:2013aca}
  M.~G.~Alford, S.~Han and M.~Prakash,
  Phys.\ Rev.\ D {\bf 88} (2013) no.8,  083013.

\bibitem{Tews:2018kmu}
  I.~Tews, J.~Carlson, S.~Gandolfi and S.~Reddy,
  Astrophys.\ J.\  {\bf 860} (2018) no.2,  149.


\bibitem{Schafer:1998ef}
  T.~Schafer and F.~Wilczek,
  Phys.\ Rev.\ Lett.\  {\bf 82} (1999) 3956.

\bibitem{Hatsuda:2006ps}
  T.~Hatsuda, M.~Tachibana, N.~Yamamoto and G.~Baym,
  Phys.\ Rev.\ Lett.\  {\bf 97} (2006) 122001.
\bibitem{Zhang:2008wx}
  Z.~Zhang, K.~Fukushima and T.~Kunihiro,
  Phys.\ Rev.\ D {\bf 79} (2009) 014004.


\bibitem{Shen:2011}
  H.~Shen, H.~Toki, K.~Oyamatsu and K.~Sumiyoshi,
  Astrophys.\ J.\ Suppl.\  {\bf 197} (2011) 20.
 

\bibitem{Oyamatsu:1993zz}
  K.~Oyamatsu,
  Nucl.\ Phys.\ A {\bf 561} (1993) 431.
 
  \bibitem{Kanzawa:2009} H. Kanzawa, M. Takano, K. Oyamatsu, K. Sumiyoshi Prog. Theor. Phys. 122, 673 (2009).
 
\bibitem{Baym:1971pw}
  G.~Baym, C.~Pethick and P.~Sutherland,
  Astrophys.\ J.\  {\bf 170} (1971) 299.
  
\bibitem{Negele:1971vb}
  J.~W.~Negele and D.~Vautherin,
  Nucl.\ Phys.\ A {\bf 207} (1973) 298.
  
\bibitem{Douchin:2001sv}
  F.~Douchin and P.~Haensel,
  Astron.\ Astrophys.\  {\bf 380} (2001) 151.
  
\bibitem{Manohar:1983md}
  A.~Manohar and H.~Georgi,
  Nucl.\ Phys.\ B {\bf 234} (1984) 189.
  
  \bibitem{Weinberg:2010bq}
  S.~Weinberg,
  Phys.\ Rev.\ Lett.\  {\bf 105} (2010) 261601.



\bibitem{Schon:2000he}
  V.~Schon and M.~Thies,
  Phys.\ Rev.\ D {\bf 62} (2000) 096002.
  
\bibitem{Bringoltz:2008iu}
  B.~Bringoltz,
  Phys.\ Rev.\ D {\bf 79} (2009) 105021.
  
\bibitem{Bringoltz:2009ym}
  B.~Bringoltz,
  Phys.\ Rev.\ D {\bf 79} (2009) 125006.

  
  \bibitem{Kojo:2011fh}
  T.~Kojo,
  Nucl.\ Phys.\ A {\bf 877} (2012) 70.
  
\bibitem{Hatsuda:1994pi}
  T.~Hatsuda and T.~Kunihiro,
  Phys.\ Rept.\  {\bf 247} (1994) 221.

 \bibitem{Buballa:2005}   M. Buballa, Phys.\ Rept.\  {\bf 407} (2005) 205.

\bibitem{Kojo:2014vja}
  T.~Kojo and G.~Baym,
  Phys.\ Rev.\ D {\bf 89} (2014) no.12,  125008.


\bibitem{Fukushima:2015bda}
  K.~Fukushima and T.~Kojo,
  Astrophys.\ J.\  {\bf 817} (2016) no.2,  180.

\bibitem{Alford:2007xm}
  M.~G.~Alford, A.~Schmitt, K.~Rajagopal and T.~Schäfer,
  Rev.\ Mod.\ Phys.\  {\bf 80} (2008) 1455.



\bibitem{Ligo-ns:2018a} Abbott, B. P.  {\it et al.} [LIGO Scientific and Virgo Collaborations] 2018, Phys. Rev. Lett. 121, 161101. 
  
\bibitem{Ligo-ns:2018b} Abbott, B.P.  {\it et al.} [LIGO Scientific and Virgo Collaborations] 2019, Phys. Rev. X  9, 011001.

\bibitem{Song:2019qoh} 
  Y.~Song, G.~Baym, T.~Hatsuda and T.~Kojo,
  arXiv:1905.01005 [astro-ph.HE].

\bibitem{Bai:2017wvk}
  Z.~Bai, H.~Chen and Y.~x.~Liu,
  Phys.\ Rev.\ D {\bf 97} (2018) no.2,  023018.

\bibitem{Zhu:2018vwn}
  Z.~Y.~Zhu, A.~Li, J.~N.~Hu and H.~Shen,
  Phys.\ Rev.\ C {\bf 99} (2019) no.2,  025804.
  
\bibitem{Ma:2018qkg}
  Y.~L.~Ma and M.~Rho,
  arXiv:1811.07071 [nucl-th].

\bibitem{McLerran:2018hbz}
  L.~McLerran and S.~Reddy,
  Phys.\ Rev.\ Lett.\  {\bf 122} (2019) no.12,  122701.


\bibitem{McLerran:2007qj}
  L.~McLerran and R.~D.~Pisarski,
  Nucl.\ Phys.\ A {\bf 796} (2007) 83.

\bibitem{Hidaka:2008yy}
  Y.~Hidaka, L.~D.~McLerran and R.~D.~Pisarski,
  Nucl.\ Phys.\ A {\bf 808} (2008) 117.

\bibitem{Kojo:2009ha}
  T.~Kojo, Y.~Hidaka, L.~McLerran and R.~D.~Pisarski,
  Nucl.\ Phys.\ A {\bf 843} (2010) 37.

\end{thebibliography}

\end{document}